\documentclass[a4paper, final, 12pt]{article}
\usepackage{eurosym}
\usepackage{amsmath, amssymb, latexsym, amscd, amsthm,amsfonts,amstext}
\usepackage[mathscr]{eucal}
\usepackage{graphicx}
\usepackage{subfig}
\usepackage{float}
\usepackage{color}
\usepackage{hyperref}
\usepackage[utf8]{inputenc}
\usepackage[english]{babel}

\numberwithin{equation}{section}
\begin{document}

\title{ Forecasting Stock Options Prices via the Solution of an Ill-Posed
Problem for the Black-Scholes Equation}
\author{Michael V. Klibanov$^{\ast }$, Aleksander A. Shananin$^{\circ }$,
Kirill V. Golubnichiy$^{\bullet }$ \\
and Sergey M. Kravchenko$^{\circ }$ \and $^{\ast }$Department of Mathematics
and Statistics, \and University of North Carolina at Charlotte, Charlotte,
NC 28223, USA \and $^{\circ }$Department of Analysis of Systems and
Solutions \and Moscow Institute of Physics and Technology, Moscow, 117303,
Russia \and $^{\bullet }$Department of Mathematics \and University of
Washington, Seattle, WA 98195, USA \and Emails: mklibanv@uncc.edu,
alexshan@yandex.ru, \and kgolubni@math.washington.edu, \and %
kravchukov1998@gmail.ru}
\date{}
\maketitle

\begin{abstract}
In the previous paper (Inverse Problems, 32, 015010, 2016), a new heuristic
mathematical model was proposed for accurate forecasting of prices of stock
options for 1-2 trading days ahead of the present one. This new technique
uses the Black-Scholes equation supplied by new intervals for the underlying
stock and \ new initial and boundary conditions for option prices. The
Black-Scholes equation was solved in the positive direction of the time
variable, This ill-posed initial boundary value problem was solved by the
so-called Quasi-Reversibility Method (QRM). This approach with an added
trading strategy was tested on the market data for 368 stock options and
good forecasting results were demonstrated. In the current paper, we use the
geometric Brownian motion to provide an explanation of that effectivity
using computationally simulated data for European call options. We also
provide a convergence analysis for QRM. The key tool of that analysis is a
Carleman estimate.
\end{abstract}

\textbf{\text{Key words: }}Black-Scholes equation, European call options, 
\text{geometric }Brownian motion, probability theory, ill-posed problem,
quasi-reversibility method, Carleman estimate, trading strategy.

\section{Introduction}

\label{sec:1}

A new heuristic mathematical algorithm designed to forecast prices of stock
options was proposed in \cite{KlibGol}. This algorithm is based on the
so-called Quasi-Reversibility Method (QRM). QRM is a regularization method
for an ill-posed problem for the Black-Scholes equation. The goal of this
paper is to address both analytically and numerically the following
question: \emph{Why this algorithm has worked well for real market data in 
\cite{KlibGol,Nik}?} Our explanations are based on our new analytical
results in the probability theory and are supported by our numerical results
for the computationally simulated data generated by the geometrical Brownian
motion.

A significant advantage of the technique of \cite{KlibGol} is that it uses
historical data about stock and option prices only over short time
intervals. This assumption is a practically valuable one since formations of
those prices are random processes. This indicates that the information used
in the algorithm possesses stable probabilistic characteristics.

The mathematical model of \cite{KlibGol} was supplied by a trading strategy.
Results of \cite[Table 4]{KlibGol} for real market data of \cite{Bloom}
indicate that a combination of that mathematical model with that trading
strategy has resulted in 72.83\% profitable options out of 368 options for
real market data. More recently, the model of \cite{KlibGol} was used in 
\cite{Nik} to forecast stock option prices in the case when results of QRM\
are enhanced by the machine learning approach, which was applied on the
second stage of the price forecasting procedure. Market data of \cite{Bloom}
for total 169,862 European call stock options were used in \cite{Nik}.
Following the machine learning approach, these data were divided in three
sets \cite[Table 1]{Nik}: training (132,912 options), validation (13,401
options) and testing (23,549 options). Total 23,549 options were tested by
QRM, and good results on predictions of options with profits were obtained
in \cite[first lines in Tables 2,3]{Nik}. Later the authors of \cite{Nik}
have tested the performance of QRM for all those 169,862 options, and
results were almost the same as ones of \cite[first lines in Tables 2,3]{Nik}%
. However, since the latter results are not yet published, then we do not
discuss them here.

\textbf{Remark 1.1: }\emph{Without further specifications, we consider in
this paper only European call options. The mathematical model of \cite%
{KlibGol} does not use neither the payment function at the expiry time nor
the strike price.}

We now present in Tables 1,2 the most recent results of \cite{Nik}, which
were obtained using the method of \cite{KlibGol} for the data consisting of
23,549 historical trades collected in 2018. The same market data of \cite%
{Bloom}\textbf{\ }were used in Tables 1,2. Option prices for one trading day
ahead of the present day were forecasted. Definitions of accuracy, precision
and recall are well known, see, e.g. \cite{Iri}.

\begin{center}
\vspace{0.5cm} \textbf{Table 1. Results of QRM for market data of \cite%
{Bloom} for 23,549 \vspace{0.5cm}options \cite[Table 2]{Nik}} \vspace{0.5cm
}

\begin{tabular}{|l|l|l|l|l|}
\hline
Method & Accuracy & Precision & Recall & Error \\ \hline
QRM & 49.77\% & 55.77\% & 52.43\% & 12 \% \\ \hline
\end{tabular}
\end{center}

In Table 1, \textquotedblleft Error" means the average relative error of
predictions of option prices, i.e.%
\begin{equation*}
\text{Error}=\frac{1}{N}\sum_{n=1}^{N}\left\vert \frac{p_{n,\text{corr}%
}-p_{n,\text{fc}}}{p_{n,\text{corr}}}\right\vert \cdot 100\%,
\end{equation*}%
where $N=23,549$ is the total number of tested options, $p_{n,\text{corr}}$
and $p_{n,\text{fc}}$ are correct and forecasted prices respectively of the
option number $n.$

\textbf{Table 2. The percentage of options with correct predictions of
profits via the Quasi-Reversibility Method for the market data of \cite%
{Bloom} for 23,549 options \cite[Table 3]{Nik}}

\begin{center}
\begin{tabular}{|l|c|l|}
\hline
Method & Correctly Predicted Profitable Options   \\ \hline
QRM & 55.77\%   \\ \hline
\end{tabular}%
\emph{\ }
\end{center}

A perfect financial market does not allow a winning strategy \cite{Fama}.
This means that to address the above question, we need to assume that the
market is imperfect. The present article considers a model situation, in
which there is a difference between the volatility $\sigma $ of the
underlying stock and traders' opinion $\hat{\sigma}$ of the volatility of an
European call option generated by this stock. We prove analytically that,
theoretically, this allows one to design a winning strategy. First, we back
up this theory numerically for the ideal case when both volatilities are
known. In practice, however, only $\hat{\sigma}$ is approximately known from 
\cite{Bloom}, where implied volatility $\sigma _{\text{impl}}$ of option
prices is posted. It is reasonable to conjecture that $\hat{\sigma}\approx
\sigma _{\text{impl}}.$

Second, to address the question posed in the first paragraph of this section
for the non ideal case, we consider a mathematical model, in which the
dynamics of the stock prices is generated by the stochastic differential
equation of the geometric Brownian motion. This allows us to computationally
generate the time series of stock prices. At the same time, we assume that
the price of the corresponding stock option is governed by the Black-Scholes
equation, in which the volatility coefficient is $\hat{\sigma}$. Hence,
using that time series of stock prices, we apply the Black-Scholes formula
to get the time series for prices of the corresponding options. Next, we
apply the QRM to predict the prices of these options for one trading day
ahead of the current one. Next, we formulate the winning strategy for the
non ideal case. 

Both the theory and the numerical studies of this paper support our two
hyphotheses formulated in subsection 6.3. Our first hypothesis that the
heuristic algorithm of \cite{KlibGol} actually figures out in many cases the
sign of the difference $\sigma -\hat{\sigma}$. Our second hypothesis is also
based on our results below as well as on the \textquotedblleft Precision"
column of Table 1 and the second column of Table 2. More precisely, the
second hypothesis is that probably about 56\% of tested 23,549 options of 
\cite{Nik} with the real market data had $\sigma -\hat{\sigma}<0.$ 

This algorithm of \cite{KlibGol} is based on the solution of a new initial
boundary value problem (IBVP) for the Black-Scholes equation, see, e.g. \cite%
{Black,W} for this equation. Since the Black-Scholes equation is actually a
1-D parabolic Partial Differential Equation (PDE) with the reversed time,
then that IBVP is ill-posed, see, e.g. \cite{KlibGol} for an example of a
high instability of a similar problem. The ill-posedness of that IBVP\ is
the main mathematical obstacle of that algorithm. Therefore, we solve that
IBVP both here and in \cite{KlibGol} by a specially designed version of QRM.
QRM stably solves this problem forwards in time for two consecutive trading
days after the current one. QRM is a version of the Tikhonov regularization
method \cite{T} being adapted to ill-posed problems for Partial Differential
Equations (PDEs). We refer to \cite{LL} for the first publication on QRM as
well as to \cite{Bourg1,Bourg2,Kl,KlibYag,KL} for some more recent ones.

We provide a convergence analysis for QRM being applied to the above
problem. The main new element of this analysis is that we lift a restrictive
assumption of \cite{KlibGol} of a sufficiently small time interval. We note
that the smallness assumption imposed on the time interval is a traditional
one for initial boundary value problems for parabolic PDEs with the reversed
time, see \cite{Kl}, \cite[Theorem 1 of section 2 in Chapter 4]{LRS}, where
a certain Carleman estimate was used. However, a new Carleman estimate was
derived in \cite{KlibYag} for a general parabolic operator of the second
order with variable coefficients in the $n-$D case. This estimate enables
one to lift that smallness assumption. We simplify here the Carleman
estimate of \cite{KlibYag} as well as some other results of \cite{KlibYag}
via adapting them to our simpler 1$-$D case, as compared with the $n-$D case
of \cite{KlibYag}. 

The Black-Scholes equation describes the dependence of the price $v\left(
s,t\right) $ of a stock option from the price of the underlying stock $s$
and time $t$ \cite{Bjork,Black,W}. In fact, this is a parabolic Partial
Differential Equation with the reversed time. Let $t=T$ be the maturity time
and $t=0$ is the present time \cite{W}. Traditionally, initial boundary
value problems for the Black-Scholes equation are solved backwards with
respect to time $t\in \left( 0,T\right) $ with the initial condition at $%
\left\{ t=T\right\} $. The latter is a well posed problem, for which the
classic theory of parabolic PDEs works, see, e.g. the book \cite{Lad} for
this theory.

However, the maturity time $T$ is usually a few months away from the present
time. It is obviously impossible to accurately predict the future behavior
of the volatility coefficient of the Black-Scholes equation on such a large
time interval. Since the formations of both stock and option prices are
stochastic processes, then it is intuitively clear a good accuracy of
forecasting of stock option prices for long time periods is unlikely.

Thus, we focus in this paper on forecasting of option prices for a short
time period of just one trading day ahead of the current one. Let the time
variable $t$ counts trading days. Since there are 255 trading days annually,
then we introduce the dimensionless time $t^{\prime }$ as%
\begin{equation}
t^{\prime }=\frac{t}{255}.  \label{1.1}
\end{equation}%
Hence, 
\begin{equation}
\text{one (1) dimensionless trading day }=1/255\approx 0.00392<<1\text{.}
\label{1.2}
\end{equation}

\textbf{Remark 1.2.} \emph{To simplify notations, we still use everywhere
below the notation }$t$\emph{\ for the dimensionless time }$t^{\prime }$%
\emph{of (\ref{1.1}).}

\textbf{Remark 1.3}. \emph{There are many important questions about the
technique of \cite{KlibGol}, which are not addressed in this paper, such as,
e.g. the performance of this technique for some \textquotedblleft stress"
tests, its performance for significantly larger sets of market data, its
performance for the case when the transaction cost is taken into account,
and many others. However, addressing any of those questions would require a
significant additional effort. Therefore, those questions are outside of the
scope of this publication. Still, the question of the transaction cost might
probably be addressed using a threshold number }$\eta >0$\emph{\ in our
trading strategy for the non-ideal case, see subsection 6.3.}

This paper is organized as follows. In section 2 we show that a winning
strategy on an infinitesimal time interval might be possible if $\sigma \neq 
\hat{\sigma}.$ In section 3 we present the heuristic mathematical model of 
\cite{KlibGol}. In section 4 we present a convergence analysis for our
version of QRM. In section 5 we use arguments of the probability theory to
justify our trading strategy in the ideal case when both volatilities $%
\sigma $ and $\hat{\sigma}$ are known. In section 6 we describe our
numerical studies and end up with a trading strategy for the non ideal case
when only the volatility $\hat{\sigma}$ is known. In addition, we formulate
in section 6 our two hyphotheses mentioned above. Concluding remarks are
given in section 7.

\textbf{Disclaimer}. \emph{This paper is written for academic purposes only.
The authors do not provide any assurance that the technique of this paper
would result in a successful trading on a real financial market.}

\section{A Possible Winning Strategy}

\label{sec:2}

Let $\sigma $ be the volatility of a certain stock and $s$ be the price of
this stock. Consider an option corresponding to this stock. Let $\hat{\sigma}
$ be an idea of the volatility of that option, which has been developed
among the agents, who trade this option on the market. If $\sigma \neq \hat{%
\sigma}$, then the financial market is imperfect, and an opportunity for
designing a winning strategy exists.

At a given time $t$, the time until the maturity will occur is $\tau ,$ 
\begin{equation}
\tau =T-t.  \label{2.0}
\end{equation}%
Let $s$ be the stock price and $f(s)$ be the payoff function of that option
at the maturity time $t=T.$ We assume that the risk-free interest rate is
zero. Let $u\left( s,\tau \right) $\ be the price of that option and the
variable $\tau $\ is the one defined in (\ref{2.0}). We assume that the
function $u\left( s,\tau \right) $ satisfies the Black-Scholes equation with
the volatility coefficient $\hat{\sigma}$ \cite[Chapter 7, Theorem 7.7]%
{Bjork}\textbf{:} 
\begin{equation}
\frac{\partial u\left( s,\tau \right) }{\partial \tau }=\frac{\hat{\sigma}%
^{2}}{2}s^{2}\frac{\partial ^{2}u\left( s,\tau \right) }{\partial s^{2}},%
\text{ }s>0,  \label{2.1}
\end{equation}%
\begin{equation*}
u\left( s,0\right) =f\left( s\right) .
\end{equation*}%
The specific formula for the payoff function is $f(s)=\max \left(
s-K,0\right) $, where $K$ is the strike price \cite{Bjork}. Then the price
function $u(s,\tau )$ of the option is given by the Black-Scholes formula 
\cite{Bjork}: 
\begin{equation}
u(s,\tau )=s\Phi (\Theta _{+}(s,\tau ))-e^{-r\tau }K\Phi (\Theta _{-}(s,\tau
)),  \label{2.2}
\end{equation}%
where $r=0$ and%
\begin{equation*}
\Theta _{\pm }\left( s,\tau \right) =\frac{1}{\hat{\sigma}\sqrt{\tau }}\left[
\ln \left( \frac{s}{K}\right) \pm \frac{\hat{\sigma}^{2}\tau }{2}\right] ,%
\text{ }
\end{equation*}%
\begin{equation}
\Phi \left( z\right) =\frac{1}{\sqrt{2\pi }}\int_{z}^{\infty }e^{-r^{2}/2}dr,%
\text{ }z\in \mathbb{R}.  \label{2.200}
\end{equation}

Let $v\left( s,t\right) =u\left( s,T-t\right) .$ The stochastic equation of
the geometric Brownian motion for the stock price $s$\ with the volatility $%
\sigma $\ has the form $ds=\sigma sdW,$\ where $W$\ is the Wiener process.
The It\^{o} formula implies

\begin{equation}
dv=\left( -\frac{\partial u(s,T-t)}{\partial \tau }+\frac{{\sigma }^{2}}{2}%
s^{2}\frac{\partial ^{2}u(s,T-t)}{\partial s^{2}}\right) dt+\sigma s\frac{%
\partial u(s,T-t)}{\partial s}dW,  \label{2.3}
\end{equation}%
where $dv$ is the option price change on an infinitesimal time interval and $%
dW$ is the Wiener process.

Replacing in (\ref{2.3}) $\partial _{\tau }u(s,T-t)$ with the right hand
side of (\ref{2.1}), we obtain

\begin{equation}
dv=\frac{(\sigma ^{2}-\hat{\sigma}^{2})}{2}s^{2}\frac{\partial ^{2}u(s,T-t)}{%
\partial s^{2}}dt+\sigma s\frac{\partial u(s,T-t)}{\partial s}dW.
\label{2.4}
\end{equation}%
The mathematical expectation of $dW$ is zero \cite[Chapter 4]{Bjork}.
Therefore, we find that the expected value of the increment of the option
price on an infinitesimal time interval is 
\begin{equation}
\frac{(\sigma ^{2}-\hat{\sigma}^{2})}{2}s^{2}\frac{\partial ^{2}u(s,T-t)}{%
\partial s^{2}}dt.  \label{2.5}
\end{equation}

In the mathematical finance, the second derivative 
\begin{equation}
\frac{\partial ^{2}u(s,\tau )}{\partial s^{2}}  \label{2.6}
\end{equation}%
is called Greek $\Gamma (s,\tau ).$ For an European call option \cite[%
Chapter 9]{Bjork} 
\begin{equation}
\Gamma (s,\tau )=\frac{1}{\hat{\sigma}s\sqrt{2\pi \tau }}\exp \left[ -\frac{%
(\Theta _{+}(s,\tau ))^{2}}{2}\right] >0.  \label{2.7}
\end{equation}

Therefore, it follows from (\ref{2.5})-(\ref{2.7}) that the sign of the
mathematical expectation of the increment of the option price on an
infinitesimal time interval is determined by the sign of the difference $%
\sigma ^{2}-\hat{\sigma}^{2}.$ Thus, if $\sigma ^{2}>\hat{\sigma}^{2},$ then
a possible winning strategy involves buying an option at the present time
and selling it in the next trading period. If $\sigma ^{2}<\hat{\sigma}^{2},$
then the winning strategy is to take the short position at the present time
and to close the short position in the next trading period.

\section{The Mathematical Model }

\label{sec:3}

\subsection{The model}

\label{sec:3.1}

We now describe the mathematical model of \cite{KlibGol}. We use this model
here for computationally simulated data. Also, it was used in \cite{Nik} for
real market data to obtain the above Tables 1,2. We do not differentiate in
this model between volatilities $\sigma $ and $\hat{\sigma}$ and just use
the time dependent volatility $\sigma \left( t\right) .$

Everywhere below, as the dimensionless time, we still use the notation $t$
for $t^{\prime }$ in (\ref{1.1}) for brevity. Let $\sigma (t)$ be the
volatility of the option at the moment of time $t$. When working with the
market data in \cite{KlibGol,Nik}, we have used the historical implied
volatility listed on the market data of \cite{Bloom}. Let $v_{b}(t)$ and $%
v_{a}(t)$ be respectively the bid and ask prices of the option and $s_{b}(t)$
and $s_{a}(t)$ be the bid and ask prices of the stock. It is known that

\begin{equation*}
v_{b}(t)<v_{a}(t)\text{ and }s_{b}(t)<s_{a}(t).
\end{equation*}%
For brevity, we simplify notations as $s_{b}=s_{b}(0)$, $s_{a}=s_{a}(0)$. We
impose a natural assumption that $0<s_{b}<s_{a}.$

It was observed on the market data in \cite[formulas (2.3)-(2.6)]{KlibGol}
that the relative differences are usually small, 
\begin{equation}
\left\vert \frac{s_{a}(t)}{s_{b}(t)}-1\right\vert \leq 0.03,\text{ }%
\left\vert \frac{v_{a}(t)}{v_{b}(t)}-1\right\vert \leq 0.27.  \label{3.0}
\end{equation}
Hence, we define the initial condition $q\left( s\right) $ at $t=0$ of the
function $v\left( s,t\right) $ as the linear interpolation on the interval $%
s\in \left( s_{b},s_{a}\right) $ between $v_{b}\left( 0\right) $ and $%
v_{a}\left( 0\right) ,$ 
\begin{equation}
v\left( s,0\right) =q\left( s\right) =-\frac{s-s_{a}}{s_{a}-s_{b}}%
v_{b}\left( 0\right) +\frac{s-s_{b}}{s_{a}-s_{b}}v_{a}\left( 0\right) ,\text{
}s\in \left( s_{b},s_{a}\right) .  \label{3.1}
\end{equation}%
Define the domain $Q_{T}=\left\{ \left( s,t\right) \in \left(
s_{b},s_{a}\right) \times \left( 0,T\right) \right\} .$ We assume that the
volatility of the option depends only on $t$, i.e. $\sigma =\sigma \left(
t\right) \geq \sigma _{0}=const.>0.$ Let $L$ be the partial differential
operator of the Black-Scholes equation, 
\begin{equation}
Lv=\frac{\partial v}{\partial t}+\frac{\sigma ^{2}\left( t\right) }{2}s^{2}%
\frac{\partial ^{2}v}{\partial s^{2}}=0\text{ in }Q_{T}.  \label{3.2}
\end{equation}%
We impose the following initial and boundary conditions on the function $%
v\left( s,t\right) :$%
\begin{equation}
v\left( s,0\right) =q\left( s\right) ,\text{ }s\in \left( s_{b},s_{a}\right)
,  \label{3.3}
\end{equation}%
\begin{equation}
v\left( s_{b},t\right) =v_{b}\left( t\right) ,v\left( s_{a},t\right)
=v_{a}\left( t\right) ,\text{ }t\in \left( 0,T\right) .  \label{3.4}
\end{equation}%
Conditions (\ref{3.2})-(\ref{3.4}) represent the heuristic mathematical
model of \cite[formulas (2.3)-(2.6)]{KlibGol}. Also, (\ref{3.2})-(\ref{3.4})
is our IBVP for the Black-Scholes equation. We now formulate this as Problem
1:

\textbf{Problem 1.} \emph{Find the function }$v\in H^{2}\left( Q_{T}\right) $%
\emph{\ satisfying conditions (\ref{3.2})-(\ref{3.4}).}

Problem 1 is ill-posed since we need to solve equation (\ref{3.2}) forwards
in time.

\textbf{Remarks 3.1:}

\begin{enumerate}
\item \textbf{\ }\emph{The conventional model for the Black-Scholes equation
stresses on the maturity time }$T$\emph{\ via considering the function }$%
u\left( s,t\right) =v\left( s,T-t\right) $\emph{\ instead of the function }$%
v\left( s,t\right) $\emph{. Unlike this, we are not doing so in (\ref{3.2})-(%
\ref{3.4}) since we do not need the maturity time, also, see Remark 1.1. }

\item \emph{As it is a conventional way in the theory of Ill-Posed problems,
we increase here the required smoothness of the solution from }$%
H^{2,1}\left( Q_{T}\right) $\emph{\ to }$H^{2}\left( Q_{T}\right) .$
\end{enumerate}

\subsection{Three steps}

\label{sec:3.2}

In order to solve Problem 1, we need first to define the time dependent
option's volatility $\sigma \left( t\right) ,$ and boundary conditions $%
v_{b}\left( t\right) $, $v_{a}\left( t\right) .$ Then the initial condition $%
q\left( s\right) $ in (\ref{3.3}) would be found via (\ref{3.3}). We explain
these in Steps 1,2 of this subsection 3.2.

In our computations of \cite{KlibGol,Nik} we have used the Implied
Volatility of the options in the Last Trade Price (IVOL) of the day for $%
\sigma \left( t\right) $ \cite{Bloom}. As to $s_{b}$ and $s_{a},$ we have
used the End of the Day Underlying Price Ask and the End of Day Underlying
Price Bid of \cite{Bloom}. Similarly for $v_{b}\left( t\right) $ and $%
v_{a}\left( t\right) ,$ in which case the End of the Day Option Price Ask
and the End of Day Option Price Bid of \cite{Bloom} were used. The moment of
time $\left\{ t=0\right\} $ is the End of the Present Day Time, and
similarly for the following two trading days of $t=y,2y$ and for the
preceding two trading days $t=-y,-2y.$ Naturally, the question can be raised
here on how did we find future values of boundary conditions $v_{b}\left(
t\right) $ and $v_{a}\left( t\right) $ for $t\in \left( 0,2y\right) $ in (%
\ref{3.4}), and the same for $\sigma \left( t\right) .$ This question is
addressed in Step 2 below. Our method for the solution of Problem 1 consists
of three steps:

\textbf{Step 1 }(introducing dimensionless variables)\textbf{. }First, we
make equation (\ref{3.2}) dimensionless. Recall that $s_{b}<s_{a}.$
Introduce the dimensionless variable $x$ for $s$ as: 
\begin{equation*}
s\Leftrightarrow x=\frac{s-s_{b}}{s_{a}-s_{b}}.
\end{equation*}

Let $y$ denotes one dimensionless trading day. By (\ref{1.2})%
\begin{equation}
y=\frac{1}{255}\approx 0.00392.  \label{3.100}
\end{equation}%
By (\ref{3.1}) the function $q\left( s\right) $ is transformed in the
function $g\left( x\right) ,$%
\begin{equation}
g\left( x\right) =\left( 1-x\right) v_{b}\left( 0\right) +xv_{a}\left(
0\right) .  \label{3.5}
\end{equation}%
And the operator $L$ in (\ref{3.2}) is transformed in the operator $M$, 
\begin{equation}
Mv=v_{t}+\sigma ^{2}\left( t\right) A(x)v_{xx},  \label{3.6}
\end{equation}%
\begin{equation}
A(x)=\frac{255}{2}\frac{[x(s_{a}-s_{b})+s_{b})]^{2}}{(s_{a}-s_{b})^{2}},
\label{3.7}
\end{equation}%
\begin{equation}
G_{2y}=\left\{ \left( x,t\right) \in \left( 0,1\right) \times \left(
0,2y\right) \right\} .  \label{3.8}
\end{equation}%
Problem is transformed in Problem 2:

\textbf{Problem 2.} \emph{Assume that functions }%
\begin{equation}
v_{b}\left( t\right) ,v_{a}\left( t\right) \in H^{2}\left[ 0,2y\right]
,\sigma \left( t\right) \in C^{1}\left[ 0,2y\right] .  \label{3.9}
\end{equation}%
\emph{Find the solution }$u\in H^{2}\left( G_{2y}\right) $\emph{\ of the
following initial boundary value problem:}%
\begin{equation}
Mv=0\text{ in }G_{2y},  \label{3.10}
\end{equation}%
\begin{equation}
v\left( 0,t\right) =v_{b}\left( t\right) ,v\left( 1,t\right) =v_{a}\left(
t\right) ,t\in \left( 0,2y\right) ,  \label{3.12}
\end{equation}%
\begin{equation}
v\left( x,0\right) =g\left( x\right) ,x\in \left( 0,1\right) ,  \label{3.11}
\end{equation}%
\emph{where the partial differential operator }$M$\emph{\ is defined in (\ref%
{3.6}), the function }$A\left( x\right) $\emph{\ is defined in (\ref{3.7}),
the initial condition }$g\left( x\right) $\emph{\ is defined in (\ref{3.5}),
and the domain }$G_{2y}$\emph{\ is defined in (\ref{3.8}).}

\textbf{Step 2 (interpolation and extrapolation).} Having the historical
market data for an option up to \textquotedblleft today", we forecast the
option price for \textquotedblleft tomorrow" and \textquotedblleft the day
after tomorrow", with 255 trading days annually. \textquotedblleft One day"
corresponds to $y=1/255.$ \textquotedblleft Today" means $t=0.$
\textquotedblleft Tomorrow" means $t=y.$ \textquotedblleft The day after
tomorrow" means $t=2y.$ We forecast these prices for the $s-$interval as $%
s\in \left[ s_{b}\left( 0\right) ,s_{a}\left( 0\right) \right] $ via the
solution of problem (\ref{3.10})-(\ref{3.12}). To do this, however, we need
to know functions $v_{b}(t),$ $v_{a}(t)$ and $\sigma (t)$ in the
\textquotedblleft future", i.e. for $t\in \left( 0,2y\right) .$ We obtain
approximate values of these functions via interpolation and extrapolation
procedures described in the next paragraph.

Let $t=-2y$ be \textquotedblleft the day before yesterday", $t=-y$ be
\textquotedblleft yesterday" and $t=0$ be \textquotedblleft today". Let $%
d\left( t\right) $ be any of three functions $v_{b}(t),v_{a}(t),\sigma (t)$.
First, we interpolate the function $d\left( t\right) $ by the quadratic
polynomial for $t\in \left[ -2y,0\right] $ using the values $d\left(
-2y\right) ,d\left( -y\right) ,d\left( 0\right) .$ We obtain%
\begin{equation}
d\left( t\right) =at^{2}+bt+c\text{ for }t\in \left[ -2y,0\right] .
\label{3.13}
\end{equation}%
Next, we extrapolate (\ref{3.13}) on the interval $t\in \left[ 0,2y\right] $
via setting%
\begin{equation*}
d\left( t\right) =at^{2}+bt+c\text{ for }t\in \left[ 0,2y\right] .
\end{equation*}%
The so defined functions $v_{b}(t),$ $v_{a}(t),\sigma (t)$ were used to
numerically solve problem (\ref{3.10})-(\ref{3.12}) for both the
computationally simulated data below and for real market data of Tables 1,2
above as well as in \cite{KlibGol}.

\textbf{Step 3 (Numerical solution of Problem 2. Regularization).} Since
problem (\ref{3.2})-(\ref{3.4}) is ill-posed, then we apply a regularization
method to obtain an approximate solution of this problem. More precisely, we
solve the following problem:

\textbf{Minimization Problem 1}. \emph{Let }$J_{\beta }:H^{2}\left(
G_{2y}\right) \rightarrow \mathbb{R}$\emph{\ be the regularization Tikhonov
functional defined as:}%
\begin{equation}
J_{\alpha }\left( v\right) =\int_{G_{2y}}\left( Mv\right) ^{2}dsdt+\alpha
\left\Vert v\right\Vert _{H^{2}\left( G_{2y}\right) }^{2},  \label{3.14}
\end{equation}%
\emph{where }$\alpha \in \left( 0,1\right) $\emph{\ is the regularization
parameter. Minimize functional (\ref{3.14}) on the set }$S,$ \emph{where} 
\begin{equation}
S=\left\{ v\in H^{2}\left( G_{2y}\right) :v\left( 0,t\right) =v_{b}\left(
t\right) ,v\left( 1,t\right) =v_{a}\left( t\right) ,v\left( x,0\right)
=g\left( x\right) \right\} .  \label{3.15}
\end{equation}

Minimization Problem 1 is a version of QRM for Problem 2. This version is an
adaptation of the QRM for problem (\ref{3.10})-(\ref{3.12}). In section 4 we
present the theory of this specific version of the QRM. In particular,
Theorem 4.2 of section 4 implies uniqueness of the solution $u\in
H^{2}\left( G_{2y}\right) $ of Problem 2 and provides an estimate of the
stability of this solution with respect to the noise in the data. Theorem
4.3 of section 4 \ implies existence and uniqueness of the minimizer $%
v_{\alpha }\in H^{2,1}\left( G_{2y}\right) $ of the functional $J_{\alpha
}\left( v\right) $ on the set $S$ defined in (\ref{3.15}). Following the
theory of Ill-Posed problems, we call such a minimizer \textquotedblleft
regularized solution" \cite{T}. Theorem 4.4 estimates convergence rate of
regularized solutions to the exact solution of Problem 2 with the noiseless
data. These estimates  depend on the noise level in the data.

\section{Convergence Analysis}

\label{sec:4}

In this section, we provide convergence analysis for Problem 2 of subsection
3.2. This problem is the IBVP for parabolic equation (\ref{3.10}) with the
reversed time, see (\ref{3.6}). The QRM for this problem for a more general
parabolic operator in $\mathbb{R}^{n}$ with arbitrary variable coefficients
was proposed in \cite{Kl} and convergence analysis was also carried out
there. Then corresponding theorems were reformulated in \cite{KlibGol}.
Although a stability estimate was not a part of \cite{KlibGol}, such an
estimate was proven in \cite{Kl}. It was pointed out in Introduction,
however, that traditional stability estimates for this problem were proven,
using a certain Carleman estimate, only under the assumption that the time
interval is sufficiently small. The same is true for the convergence
theorems of QRM\ in \cite{Kl,KlibGol}. Unlike this, the smallness assumption
was lifted in \cite{KlibYag} via a new Carleman estimate. In this section,
we significantly modify results of \cite{KlibYag} for a simpler 1-D case.
Recall (see Introduction) that this modification allows us to obtain more
accurate estimates in the 1-D case, as compared with the $n-$D case of \cite%
{KlibYag}. We note that even though we work in our computations below on a
small time interval $\left( 0,2y\right) =\left( 0,0.00784\right) $ (see (\ref%
{3.100}) and (\ref{3.8})), the smallness assumption of \cite{Kl,KlibGol}, 
\cite[Theorem 1 of section 2 in Chapter 4]{LRS} might result in the
requirement of even a smaller length of that interval.

\subsection{Problem statement}

\label{sec:4.1}

Consider a number $T_{1}>0$ and denote%
\begin{equation*}
Q_{T_{1}}=\left\{ \left( x,t\right) \in \left( 0,1\right) \times \left(
0,T_{1}\right) \right\} .
\end{equation*}%
Let two numbers $a_{0},a_{1}>0$ and $a_{0}<a_{1}.$ Let the function $a\left(
x,t\right) \in C^{1}\left( \overline{Q}_{T_{1}}\right) $ satisfies:%
\begin{equation}
\text{ }\left\Vert a\right\Vert _{C^{1}\left( \overline{Q}_{T_{1}}\right)
}\leq a_{1},\text{ }a\left( x,t\right) \geq a_{0}\text{ in }Q_{T_{1}}.
\label{7.1}
\end{equation}%
Let functions $\varphi _{0}\left( t\right) ,\varphi _{1}\left( t\right) \in
H^{2}\left( 0,T_{1}\right) .$ In the above case of subsection 3.2, 
\begin{equation*}
T_{1}=2y,a\left( x,t\right) =\sigma ^{2}\left( t\right) A(x),\varphi
_{0}\left( t\right) =v_{b}\left( t\right) ,\varphi _{1}\left( t\right)
=v_{a}\left( t\right) .
\end{equation*}%
We now formulate Problem 3, which is a slight generalization of Problem 2.

\textbf{Problem 3}. \emph{Find a solution }$w\in H^{2}\left(
Q_{T_{1}}\right) $\emph{\ of the following initial boundary value problem
(IBVP):}%
\begin{equation}
Pw=w_{t}+a\left( x,t\right) w_{xx}=0\text{ in }Q_{T_{1}},  \label{7.3}
\end{equation}%
\begin{equation}
w\left( 0,t\right) =\varphi _{0}\left( t\right) ,w\left( 1,t\right) =\varphi
_{1}\left( t\right) ,\text{ }t\in \left( 0,T_{1}\right) ,  \label{7.4}
\end{equation}%
\begin{equation}
w\left( x,0\right) =q\left( x\right) =\varphi _{0}\left( 0\right) \left(
1-x\right) +\varphi _{1}\left( 0\right) x,\text{ }x\in \left( 0,1\right) .
\label{7.5}
\end{equation}

\textbf{Remark 4.1.} \emph{Since Problem 3 is a more general one than
Problem 2, then our convergence analysis for Problem 3, which we provide
below, is also valid for Problem 2.}

The reason why we use the linear function for $w\left( x,0\right) $ in (\ref%
{7.5}) is our desire to simplify the presentation by using the fact that, in
the case of Problem 2, the initial condition in (\ref{3.11}) is the linear
function defined in (\ref{3.5}). Problem 3 is an IBVP\ for the parabolic
equation (\ref{7.3}) with the reversed time. Therefore, this problem is
ill-posed. Just as it is always the case in the theory of Ill-Posed problems 
\cite{T}, we assume that the boundary in (\ref{7.4}) are given with a noise
of the level $\delta >0,$ where $\delta $ is a sufficiently small number,
i.e.%
\begin{equation}
\left\Vert \varphi _{0}-\varphi _{0}^{\ast }\right\Vert _{H^{1}\left(
0,T_{1}\right) }<\delta ,\left\Vert \varphi _{1}-\varphi _{1}^{\ast
}\right\Vert _{H^{1}\left( 0,T_{1}\right) }<\delta ,  \label{7.6}
\end{equation}%
where functions $\varphi _{0}^{\ast },\varphi _{1}^{\ast }\in H^{2}\left(
0,T_{1}\right) $ are \textquotedblleft ideal" noiseless data. Following to
one of postulates of the theory of Ill-Posed problems, we assume that there
exists an exact solution $w^{\ast }\in H^{2}\left( Q_{T_{1}}\right) $ of
problem (\ref{7.3})-(\ref{7.5}) with these noiseless data. We will estimate
below how this noise affects the accuracy of the solution of Problem 3 (if
this solution exists) and also will establish the convergence rate of
numerical solutions obtained by QRM to the exact one as $\delta \rightarrow
0.$

Consider the following analog of functional (\ref{3.14}):%
\begin{equation}
I_{\alpha }\left( w\right) =\int_{Q_{T_{1}}}\left( Pw\right) ^{2}dxdt+\alpha
\left\Vert w\right\Vert _{H^{2}\left( Q_{T_{1}}\right) }^{2}.  \label{7.7}
\end{equation}%
Introduce the set $Y\subset H^{2}\left( Q_{T_{1}}\right) ,$%
\begin{equation}
Y=\left\{ w\in H^{2}\left( Q_{T_{1}}\right) :w\left( 0,t\right) =\varphi
_{0}\left( t\right) ,w\left( 1,t\right) =\varphi _{1}\left( t\right)
,w\left( x,0\right) =q\left( x\right) \right\} .  \label{7.8}
\end{equation}%
We construct an approximate solution of Problem 3 via solving the following
problem:

\textbf{Minimization Problem 2}. \emph{Minimize the functional }$I_{\alpha
}\left( w\right) $\emph{\ on the set }$Y$\emph{\ given in (\ref{7.8}).}

Similarly with the Minimization Problem1, Minimization Problem 2 means QRM\
for Problem 3.

\subsection{Theorems}

\label{sec:4.2}

In this subsection, we formulate four theorems for Problem 3. Let $\lambda
>2 $ be a parameter. Introduce the Carleman Weight Function $\psi _{\lambda
}\left( t\right) $ for the operator $\partial _{t}+a\left( x,t\right)
\partial _{x}^{2}$ as:%
\begin{equation}
\psi _{\lambda }\left( t\right) =e^{\left( T_{1}+1-t\right) ^{\lambda }},%
\text{ }t\in \left( 0,T_{1}\right) .  \label{7.9}
\end{equation}%
Hence, the function $\psi _{\lambda }\left( t\right) $ is decreasing on $%
\left[ 0,T_{1}\right] $, $\psi _{\lambda }^{\prime }\left( t\right) <0,$%
\begin{equation}
\max_{\left[ 0,T_{1}\right] }\psi _{\lambda }\left( t\right) =\psi _{\lambda
}\left( 0\right) =e^{\left( T_{1}+1\right) ^{\lambda }},\text{ }\min_{\left[
0,T_{1}\right] }\psi _{\lambda }\left( t\right) =\psi _{\lambda }\left(
T_{1}\right) =e.  \label{7.10}
\end{equation}%
Denote%
\begin{equation}
H_{0}^{2}\left( Q_{T_{1}}\right) =\left\{ u\in H^{2}\left( Q_{T_{1}}\right)
:u\left( 0,t\right) =u\left( 1,t\right) =0\right\} .  \label{7.99}
\end{equation}%
\begin{equation}
H_{0,0}^{2}\left( Q_{T_{1}}\right) =\left\{ u\in H_{0}^{2}\left(
Q_{T_{1}}\right) :u\left( x,0\right) =0\right\} .  \label{7.100}
\end{equation}

\textbf{Theorem 4.1} (Carleman estimate). \emph{Let the coefficient }$%
a\left( x,t\right) $\emph{\ of the operator }$P$\emph{\ satisfies conditions
(\ref{7.1}). Then there exist a sufficiently large number }$\lambda
_{0}=\lambda _{0}\left( T_{1},a_{0},a_{1}\right) >2$\emph{\ and a constant }$%
C=C\left( T_{1},a_{0},a_{1}\right) >0,$\emph{\ both depending only on listed
parameters, such that the following Carleman estimate holds for the operator 
}$P:$\emph{\ }%
\begin{equation*}
\int_{Q_{T_{1}}}\left( Pu\right) ^{2}\psi _{\lambda }^{2}dxdt\geq C\sqrt{%
\lambda }\int_{Q_{T_{1}}}u_{x}^{2}\psi _{\lambda }^{2}dxdt+C\lambda
^{2}\int_{Q_{T_{1}}}u^{2}\psi _{\lambda }^{2}dxdt
\end{equation*}%
\begin{equation}
-C\sqrt{\lambda }\left\Vert u\right\Vert _{H^{2}\left( Q_{T_{1}}\right)
}^{2}-C\lambda \left( T_{1}+1\right) ^{\lambda }e^{2\left( T_{1}+1\right)
^{\lambda }}\left\Vert u\left( x,0\right) \right\Vert _{L_{2}\left(
0,1\right) }^{2},  \label{7.11}
\end{equation}%
\begin{equation*}
\forall \lambda \geq \lambda _{0},\forall u\in H_{0}^{2}\left(
Q_{T_{1}}\right) .
\end{equation*}

Carleman estimate (\ref{7.11}) is the key to proofs of Theorems 4.2, 4.4.

\textbf{Theorem 4.2} (H\"{o}lder stability estimate for Problem 3 and
uniqueness). \emph{Let the coefficient }$a\left( x,t\right) $\emph{\ of the
operator }$P$\emph{\ satisfies conditions (\ref{7.1}). Assume that the
functions }$w\in H^{2}\left( Q_{T_{1}}\right) $\emph{\ and }$w^{\ast }\in
H^{2}\left( Q_{T_{1}}\right) $\emph{\ are solutions of Problem 3 with the
vectors of data }$\left( \varphi _{0}\left( t\right) ,\varphi _{1}\left(
t\right) \right) $\emph{\ and }$\left( \varphi _{0}^{\ast }\left( t\right)
,\varphi _{1}^{\ast }\left( t\right) \right) $\emph{\ respectively, where }$%
\varphi _{0},\varphi _{1},\varphi _{0}^{\ast },\varphi _{1}^{\ast }\in
H^{2}\left( 0,T_{1}\right) .$\emph{\ Also, assume that error estimates (\ref%
{7.6}) of the boundary data hold. Choose an arbitrary number }$\rho \in
\left( 0,T_{1}\right) $\emph{. Denote }%
\begin{equation}
\mu =\mu \left( T_{1},\rho \right) =\frac{\ln \left( T_{1}+1-\rho \right) }{%
\ln \left( T_{1}+1\right) }\in \left( 0,1\right) .  \label{7.110}
\end{equation}%
\emph{Then there exists a sufficiently small number }$\delta _{0}=\delta
_{0}\left( T_{1},a_{0},a_{1}\right) \in \left( 0,1\right) $\emph{\ and a
constant }$C_{1}=C_{1}\left( T_{1},a_{0},a_{1},\rho \right) >0,$ \emph{both
depending only on listed parameters, such that\ the following stability
estimate holds for all }$\delta \in \left( 0,\delta _{0}\right) :$%
\begin{equation}
\left\Vert w_{x}-w_{x}^{\ast }\right\Vert _{L_{2}\left( Q_{T_{1}-\rho
}\right) }+\left\Vert w-w^{\ast }\right\Vert _{L_{2}\left( Q_{T_{1}-\rho
}\right) }\leq  \label{7.12}
\end{equation}%
\begin{equation*}
\leq C_{1}\left( 1+\left\Vert w-w^{\ast }\right\Vert _{H^{2}\left(
Q_{T_{1}}\right) }\right) \exp \left[ -\left( \ln \delta ^{-1/2}\right)
^{\mu }\right] .
\end{equation*}

Below $C=C\left( T_{1},a_{0},a_{1}\right) >0$ and $C_{1}=C_{1}\left(
T_{1},a_{0},a_{1}\right) >0$ denote different constants depending only on
listed parameters.

\textbf{Corollary 4.1} (uniqueness). \emph{Let the coefficient }$a\left(
x,t\right) $\emph{\ of the operator }$P$\emph{\ satisfies conditions (\ref%
{7.1}). Then Problem 3 has at most one solution (uniqueness)}.

\textbf{Proof}.\emph{\ }If $\delta =0,$\ then (\ref{7.12}) implies that $%
w\left( x,t\right) =w^{\ast }\left( x,t\right) $\ in $Q_{T_{1}-\rho }.$
Since $\rho \in \left( 0,T_{1}\right) $\ is an arbitrary number, then $%
w\left( x,t\right) \equiv w^{\ast }\left( x,t\right) $\ in $Q_{T_{1}}.$ $%
\square $

\textbf{Theorem 4.3 }(existence and uniqueness of the minimizer)\textbf{. }%
\emph{Let functions }$\varphi _{0}\left( t\right) ,\varphi _{1}\left(
t\right) \in H^{2}\left( 0,T_{1}\right) .$\emph{\ Let }$Y$\emph{\ be the set
defined in (\ref{7.8}). Then there exists unique minimizer }$w_{\min }\in Y$%
\emph{\ of functional (\ref{7.7}) and }%
\begin{equation}
\left\Vert w_{\min }\right\Vert _{H^{2}\left( Q_{T_{1}}\right) }\leq \frac{C%
}{\sqrt{\alpha }}\left( \left\Vert \varphi _{0}\right\Vert _{H^{2}\left(
0,T_{1}\right) }+\left\Vert \varphi _{1}\right\Vert _{H^{2}\left(
0,T_{1}\right) }\right) .  \label{7.13}
\end{equation}

In the theory of Ill-Posed Problems, this minimizer $w_{\min }$ is called
\textquotedblleft regularized solution" of Problem 3 \cite{T}. According to
the theory of Ill-Posed problems, it is important to establish convergence
rate of regularized solutions to the exact one $w^{\ast }.$ In doing so, one
should always choose a dependence of the regularization parameter $\alpha $
on the noise level $\delta ,$ i.e. $\alpha =\alpha \left( \delta \right) \in
\left( 0,1\right) $ \cite{T}.

\textbf{Theorem 4.4} (convergence rate of regularized solutions). \emph{Let }%
$w^{\ast }\in H^{2}\left( Q_{T_{1}}\right) $\emph{\ be the solution of
Problem 3 with the noiseless data }$\left( \varphi _{0}^{\ast }\left(
t\right) ,\varphi _{1}^{\ast }\left( t\right) \right) .$\emph{\ Let
functions }$\varphi _{0},\varphi _{1},\varphi _{0}^{\ast },\varphi
_{1}^{\ast }\in H^{2}\left( 0,T_{1}\right) .$\emph{\ Let }$w_{\min }\in Y$%
\emph{\ be the unique minimizer of functional (\ref{7.7}) on the set }$Y$%
\emph{. Assume that error estimates (\ref{7.6}) hold. Choose an arbitrary
number }$\rho \in \left( 0,T_{1}\right) $\emph{. Let }$\mu =\mu \left(
T_{1},\rho \right) \in \left( 0,1\right) $ \emph{be the number defined in (%
\ref{7.110}) and let}%
\begin{equation}
\alpha =\alpha \left( \delta \right) =\delta ^{2},  \label{7.140}
\end{equation}%
\emph{\ Then there exists a sufficiently small number }$\delta _{0}=\delta
_{0}\left( T_{1},a_{0},a_{1}\right) \in \left( 0,1\right) $\emph{\ depending
only on listed parameters such that the following convergence rate of
regularized solutions }$w_{\min }$ \emph{holds for all }$\delta \in \left(
0,\delta _{0}\right) :$\emph{\ }%
\begin{equation}
\left\Vert \partial _{x}w_{\min }-\partial _{x}w^{\ast }\right\Vert
_{L_{2}\left( Q_{T_{1}-\rho }\right) }+\left\Vert w_{\min }-w^{\ast
}\right\Vert _{L_{2}\left( Q_{T_{1}-\rho }\right) }  \label{7.14}
\end{equation}%
\begin{equation*}
\leq C_{1}\left( 1+\left\Vert w^{\ast }\right\Vert _{H^{2}\left(
Q_{T_{1}}\right) }+\left\Vert \varphi _{0}^{\ast }\right\Vert _{H^{2}\left(
0,T_{1}\right) }+\left\Vert \varphi _{1}^{\ast }\right\Vert _{H^{2}\left(
0,T_{1}\right) }\right) \exp \left[ -\left( \ln \delta ^{-1/2}\right) ^{\mu }%
\right] .
\end{equation*}

\subsection{Proof of Theorem 4.1}

\label{sec:4.3}

We assume in this proof that $u\in C^{2}\left( \overline{Q}_{T_{1}}\right) .$
The case $u\in H^{2}\left( Q_{T_{1}}\right) $ can be obtained via density
arguments. It is assumed in this proof that $\lambda \geq \lambda
_{0}=\lambda _{0}\left( T_{1},a_{0},a_{1}\right) >2$ and $\lambda _{0}$ is
sufficiently large. We remind that $C=C\left( T_{1},a_{0},a_{1}\right) >0$
denotes different constants depending only on listed parameters. Change
variables as 
\begin{equation}
v\left( x,t\right) =u\left( x,t\right) \psi _{\lambda }\left( t\right)
=u\left( x,t\right) e^{\left( T_{1}+1-t\right) ^{\lambda }}.  \label{7.16}
\end{equation}%
Hence, 
\begin{equation*}
u\left( x,t\right) =v\left( x,t\right) e^{-\left( T_{1}+1-t\right) ^{\lambda
}},
\end{equation*}%
\begin{equation*}
u_{t}=\left( v_{t}+\lambda \left( T_{1}+1-t\right) ^{\lambda -1}v\right)
e^{-\left( T_{1}+1-t\right) ^{\lambda }},\text{ }
\end{equation*}%
\begin{equation*}
u_{x}=v_{x}e^{-\left( T_{1}+1-t\right) ^{\lambda }},\text{ }%
u_{xx}=v_{xx}e^{-\left( T_{1}+1-t\right) ^{\lambda }}.
\end{equation*}%
Hence, 
\begin{equation*}
\left( Pu\right) ^{2}\psi _{\lambda }^{2}=\left[ v_{t}+\left( a\left(
x,t\right) v_{xx}+\lambda \left( T_{1}+1-t\right) ^{\lambda -1}v\right) %
\right] ^{2}
\end{equation*}%
\begin{equation}
\geq v_{t}^{2}+2v_{t}\left( a\left( x,t\right) v_{xx}+\lambda \left(
T_{1}+1-t\right) ^{\lambda -1}v\right) .  \label{7.17}
\end{equation}%
We have used here $\left( a+b\right) ^{2}\geq a^{2}+2ab,$ $\forall a,b\in 
\mathbb{R}.$ We now estimate from the below terms in the second line of (\ref%
{7.17}).

\textbf{Step 1}. Estimate from the below $2a\left( x,t\right) v_{xx}v_{t}.$
We have:%
\begin{equation*}
2a\left( x,t\right) v_{xx}v_{t}=\left( 2a\left( x,t\right) v_{x}v_{t}\right)
_{x}-2a\left( x,t\right) v_{x}v_{xt}-2a_{x}\left( x,t\right) v_{x}v_{t}
\end{equation*}%
\begin{equation*}
=\left( 2a\left( x,t\right) v_{x}v_{t}\right) _{x}+\left( -a\left(
x,t\right) v_{x}^{2}\right) _{t}-a_{t}\left( x,t\right)
v_{x}^{2}-2a_{x}\left( x,t\right) v_{x}v_{t}.
\end{equation*}%
Thus, 
\begin{equation}
2a\left( x,t\right) v_{xx}v_{t}\geq \left( 2a\left( x,t\right)
v_{x}v_{t}\right) _{x}+\left( -a\left( x,t\right) v_{x}^{2}\right)
_{t}-Cv_{x}^{2}-C\left\vert v_{x}\right\vert \left\vert v_{t}\right\vert .
\label{7.18}
\end{equation}

\textbf{Step 2.} Estimate from the below $2\lambda \left( T_{1}+1-t\right)
^{\lambda -1}vv_{t}.$ We have:%
\begin{equation*}
2\lambda \left( T_{1}+1-t\right) ^{\lambda -1}vv_{t}=\left( \lambda \left(
T_{1}+1-t\right) ^{\lambda -1}v^{2}\right) _{t}+\lambda \left( \lambda
-1\right) \left( T_{1}+1-t\right) ^{\lambda -2}v^{2}
\end{equation*}%
\begin{equation}
\geq \left( \lambda \left( T_{1}+1-t\right) ^{\lambda -1}v^{2}\right) _{t}+%
\frac{\lambda ^{2}}{2}\left( T_{1}+1-t\right) ^{\lambda -2}v^{2}.
\label{7.19}
\end{equation}

\textbf{Step 3}. Estimate from the below the entire second line of (\ref%
{7.17}). Using (\ref{7.18}), (\ref{7.19}) and Cauchy-Schwarz inequality
\textquotedblleft with $\varepsilon ",$%
\begin{equation}
2ab\geq -\varepsilon a^{2}-\frac{1}{\varepsilon }b^{2},\text{ }\forall
a,b\in \mathbb{R},\text{ }\forall \varepsilon >0,  \label{7.190}
\end{equation}%
we obtain%
\begin{equation*}
v_{t}^{2}+2v_{t}\left( a\left( x,t\right) v_{xx}+\lambda \left(
T_{1}+1-t\right) ^{\lambda -1}v\right) \geq
\end{equation*}%
\begin{equation*}
\geq v_{t}^{2}-Cv_{x}^{2}-C\left\vert v_{x}\right\vert \left\vert
v_{t}\right\vert +\frac{\lambda ^{2}}{2}\left( T_{1}+1-t\right) ^{\lambda
-2}v^{2}
\end{equation*}%
\begin{equation*}
+\left( 2a\left( x,t\right) v_{x}v_{t}\right) _{x}+\left( -a\left(
x,t\right) v_{x}^{2}+\lambda \left( T_{1}+1-t\right) ^{\lambda
-1}v^{2}\right) _{t}
\end{equation*}%
\begin{equation*}
\geq \frac{1}{2}v_{t}^{2}-Cv_{x}^{2}+\frac{\lambda ^{2}}{2}\left(
T_{1}+1-t\right) ^{\lambda -2}v^{2}
\end{equation*}%
\begin{equation*}
+\left( 2a\left( x,t\right) v_{x}v_{t}\right) _{x}+\left( -a\left(
x,t\right) v_{x}^{2}+\lambda \left( T_{1}+1-t\right) ^{\lambda
-1}v^{2}\right) _{t}.
\end{equation*}%
Thus, we have obtained that 
\begin{equation*}
v_{t}^{2}+2v_{t}\left( a\left( x,t\right) v_{xx}+\lambda \left(
T_{1}+1-t\right) ^{\lambda -1}v\right) \geq
\end{equation*}%
\begin{equation}
\geq \frac{1}{2}v_{t}^{2}-Cv_{x}^{2}+\frac{\lambda ^{2}}{2}\left(
T_{1}+1-t\right) ^{\lambda -2}v^{2}  \label{7.20}
\end{equation}%
\begin{equation*}
+\left( 2a\left( x,t\right) v_{x}v_{t}\right) _{x}+\left( -a\left(
x,t\right) v_{x}^{2}+\lambda \left( T_{1}+1-t\right) ^{\lambda
-1}v^{2}\right) _{t}.
\end{equation*}%
Using (\ref{7.17}) and (\ref{7.20}) as well as dropping the non-negative
term $v_{t}^{2}/2$ in the right hand side of (\ref{7.20}), we obtain%
\begin{equation}
\left( Pu\right) ^{2}\psi _{\lambda }^{2}\geq -Cv_{x}^{2}+\frac{\lambda ^{2}%
}{2}\left( T_{1}+1-t\right) ^{\lambda -2}v^{2}  \label{7.21}
\end{equation}%
\begin{equation*}
+\left( 2a\left( x,t\right) v_{x}v_{t}\right) _{x}+\left( -a\left(
x,t\right) v_{x}^{2}+\lambda \left( T_{1}+1-t\right) ^{\lambda
-1}v^{2}\right) _{t}.
\end{equation*}

\textbf{Step 4}. Using (\ref{7.16}), change variables in the right hand side
of (\ref{7.21}). We have $v^{2}=u^{2}\psi _{\lambda
}^{2},v_{x}^{2}=u_{x}^{2}\psi _{\lambda }^{2}.$ Thus,%
\begin{equation}
\left( Pu\right) ^{2}\psi _{\lambda }^{2}\geq -Cu_{x}^{2}\psi _{\lambda
}^{2}+\frac{\lambda ^{2}}{2}\left( T_{1}+1-t\right) ^{\lambda -2}u^{2}\psi
_{\lambda }^{2}  \label{7.22}
\end{equation}%
\begin{equation*}
+\left( 2a\left( x,t\right) u_{x}\left( u_{t}-\lambda \left(
T_{1}+1-t\right) ^{\lambda -2}u\right) \psi _{\lambda }^{2}\right)
_{x}+\left( \left( -a\left( x,t\right) u_{x}^{2}+\lambda \left(
T_{1}+1-t\right) ^{\lambda -1}u^{2}\right) \psi _{\lambda }^{2}\right) _{t}.
\end{equation*}

\textbf{Step 5.} Estimate from the below $-Pu\cdot u\psi _{\lambda }^{2}.$
We have 
\begin{equation*}
-Pu\cdot u\psi _{\lambda }^{2}=\left( -u_{t}-a\left( x,t\right)
u_{xx}\right) ue^{2\left( T_{1}+1-t\right) ^{\lambda }}
\end{equation*}%
\begin{equation}
=\left( -\frac{1}{2}u^{2}e^{2\left( T_{1}+1-t\right) ^{\lambda }}\right)
_{t}-\lambda \left( T_{1}+1-t\right) ^{\lambda -1}u^{2}e^{2\left(
T_{1}+1-t\right) ^{\lambda }}  \label{7.23}
\end{equation}%
\begin{equation*}
+\left( -a\left( x,t\right) u_{x}ue^{2\left( T_{1}+1-t\right) ^{\lambda
}}\right) _{x}+a\left( x,t\right) u_{x}^{2}e^{2\left( T_{1}+1-t\right)
^{\lambda }}+a_{x}\left( x,t\right) u_{x}ue^{2\left( T_{1}+1-t\right)
^{\lambda }}.
\end{equation*}%
Using (\ref{7.1}) and (\ref{7.190}), we obtain%
\begin{equation*}
a\left( x,t\right) u_{x}^{2}+a_{x}\left( x,t\right) u_{x}u\geq
a_{0}u_{x}^{2}-a_{1}\left\vert u_{x}\right\vert \left\vert u\right\vert \geq 
\frac{a_{0}}{2}u_{x}^{2}-Cu^{2}
\end{equation*}%
\begin{equation*}
\geq \frac{a_{0}}{2}u_{x}^{2}-\lambda \left( T_{1}+1-t\right) ^{\lambda
-2}u^{2}.
\end{equation*}%
Hence, multiplying (\ref{7.23}) by $\sqrt{\lambda }$, we obtain%
\begin{equation}
-\sqrt{\lambda }Pu\cdot u\psi _{\lambda }^{2}\geq \frac{a_{0}}{2}\sqrt{%
\lambda }u_{x}^{2}e^{2\left( T_{1}+1-t\right) ^{\lambda }}-2\lambda
^{3/2}\left( T_{1}+1-t\right) ^{\lambda -2}u^{2}e^{2\left( T_{1}+1-t\right)
^{\lambda }}  \label{7.24}
\end{equation}%
\begin{equation*}
+\left( -\frac{\sqrt{\lambda }}{2}u^{2}e^{2\left( T_{1}+1-t\right) ^{\lambda
}}\right) _{t}+\left( -\sqrt{\lambda }a\left( x,t\right) u_{x}ue^{2\left(
T_{1}+1-t\right) ^{\lambda }}\right) _{x}.
\end{equation*}

\textbf{Step 6}. Estimate from the below $\left( Pu\right) ^{2}\psi
_{\lambda }^{2}-\sqrt{\lambda }Pu\cdot u\psi _{\lambda }^{2}.$ Using (\ref%
{7.22}) and (\ref{7.24}), we obtain%
\begin{equation*}
\left( Pu\right) ^{2}\psi _{\lambda }^{2}-\sqrt{\lambda }Pu\cdot u\psi
_{\lambda }^{2}\geq
\end{equation*}%
\begin{equation*}
\geq \frac{a_{0}}{2}\sqrt{\lambda }\left( 1-\frac{2C}{\sqrt{\lambda }}%
\right) u_{x}^{2}\psi _{\lambda }^{2}+\frac{\lambda ^{2}}{2}\left(
T_{1}+1-t\right) ^{\lambda -2}\left( 1-\frac{4}{\sqrt{\lambda }}\right)
u^{2}\psi _{\lambda }^{2}
\end{equation*}%
\begin{equation}
+\frac{\partial }{\partial t}\left[ \left( -a\left( x,t\right)
u_{x}^{2}+\lambda \left( T_{1}+1-t\right) ^{\lambda -1}u^{2}-\frac{\sqrt{%
\lambda }}{2}u^{2}\right) \psi _{\lambda }^{2}\right]  \label{7.25}
\end{equation}%
\begin{equation*}
+\frac{\partial }{\partial x}\left[ \left( 2a\left( x,t\right) u_{x}\left(
u_{t}-\lambda \left( T_{1}+1-t\right) ^{\lambda -2}u\right) -\sqrt{\lambda }%
a\left( x,t\right) u_{x}u\right) \psi _{\lambda }^{2}\right] .
\end{equation*}

\textbf{Step 7}. Estimate from the below 
\begin{equation*}
\int_{Q_{T_{1}}}\left( Pu\right) ^{2}\psi _{\lambda }^{2}dxdt.
\end{equation*}%
We have 
\begin{equation*}
\left( Pu\right) ^{2}\psi _{\lambda }^{2}-\sqrt{\lambda }Pu\cdot u\psi
_{\lambda }^{2}\leq \frac{3}{2}\left( Pu\right) ^{2}\psi _{\lambda }^{2}+%
\frac{1}{2}\sqrt{\lambda }u^{2}\psi _{\lambda }^{2}.
\end{equation*}%
Combining this with (\ref{7.25}), we obtain%
\begin{equation*}
\left( Pu\right) ^{2}\psi _{\lambda }^{2}\geq C\sqrt{\lambda }u_{x}^{2}\psi
_{\lambda }^{2}+C\lambda ^{2}u^{2}\psi _{\lambda }^{2}
\end{equation*}%
\begin{equation}
+\frac{\partial }{\partial t}\left[ \left( -a\left( x,t\right)
u_{x}^{2}+\lambda \left( T_{1}+1-t\right) ^{\lambda -1}u^{2}-\frac{\sqrt{%
\lambda }}{2}u^{2}\right) \psi _{\lambda }^{2}\right]  \label{7.26}
\end{equation}%
\begin{equation*}
+\frac{\partial }{\partial x}\left[ \left( 2a\left( x,t\right) u_{x}\left(
u_{t}-\lambda \left( T_{1}+1-t\right) ^{\lambda -2}u\right) -\sqrt{\lambda }%
a\left( x,t\right) u_{x}u\right) \psi _{\lambda }^{2}\right] .
\end{equation*}%
Integrate (\ref{7.26}) using $u\in H_{0}^{2}\left( Q_{T_{1}}\right) $ and
also using (\ref{7.10}). We obtain 
\begin{equation*}
\int_{Q_{T_{1}}}\left( Pu\right) ^{2}\psi _{\lambda }^{2}dxdt\geq C\sqrt{%
\lambda }\int_{Q_{T_{1}}}u_{x}^{2}\psi _{\lambda }^{2}dxdt+C\lambda
^{2}\int_{Q_{T_{1}}}u^{2}\psi _{\lambda }^{2}dxdt
\end{equation*}%
\begin{equation}
-C\sqrt{\lambda }\left\Vert u\left( x,T_{1}\right) \right\Vert _{H^{1}\left(
0,1\right) }^{2}-C\lambda \left( T_{1}+1\right) ^{\lambda }e^{2\left(
T_{1}+1\right) ^{\lambda }}\left\Vert u\left( x,0\right) \right\Vert
_{L_{2}\left( 0,1\right) }^{2}.  \label{7.27}
\end{equation}%
Finally, applying the trace theorem to the second line of (\ref{7.27}), we
obtain desired estimate (\ref{7.11}) of this theorem. $\square $

\subsection{Proof of Theorem 4.2}

\label{sec:4.4}

Introduce the following functions:%
\begin{equation}
\widetilde{\varphi }_{0}\left( t\right) =\varphi _{0}\left( t\right)
-\varphi _{0}^{\ast }\left( t\right) ,\widetilde{\varphi }_{1}\left(
t\right) =\varphi _{1}\left( t\right) -\varphi _{1}^{\ast }\left( t\right) ,
\label{7.28}
\end{equation}%
\begin{equation}
F\left( x,t\right) =\varphi _{0}\left( t\right) \left( 1-x\right) +\varphi
_{1}\left( t\right) x,\text{ }F^{\ast }\left( x,t\right) =\varphi _{0}^{\ast
}\left( t\right) \left( 1-x\right) +\varphi _{1}^{\ast }\left( t\right) x,
\label{7.29}
\end{equation}%
\begin{equation}
\widetilde{F}\left( x,t\right) =F\left( x,t\right) -\text{ }F^{\ast }\left(
x,t\right) =\widetilde{\varphi }_{0}\left( t\right) \left( 1-x\right) +%
\widetilde{\varphi }_{1}\left( t\right) ,  \label{7.30}
\end{equation}%
\begin{equation}
\widehat{w}\left( x,t\right) =w\left( x,t\right) -F\left( x,t\right) ,%
\widehat{w}^{\ast }\left( x,t\right) =w^{\ast }\left( x,t\right) -F^{\ast
}\left( x,t\right) ,  \label{7.31}
\end{equation}%
\begin{equation}
\overline{w}\left( x,t\right) =\widehat{w}\left( x,t\right) -\widehat{w}%
^{\ast }\left( x,t\right) .  \label{7.32}
\end{equation}%
It follows from (\ref{7.5}), (\ref{7.6}) and (\ref{7.28})-(\ref{7.32}) that:%
\begin{equation}
\widehat{w}\left( x,0\right) =\widehat{w}^{\ast }\left( x,0\right) =%
\overline{w}\left( x,0\right) =0,  \label{7.34}
\end{equation}%
\begin{equation}
F_{xx}\left( x,t\right) =\text{ }F_{xx}^{\ast }\left( x,t\right) =\widetilde{%
F}_{xx}\left( x,t\right) =0,  \label{7.340}
\end{equation}%
\begin{equation}
\left\Vert \widetilde{F}_{t}\right\Vert _{L_{2}\left( Q_{T_{1}}\right)
},\left\Vert \widetilde{F}\right\Vert _{L_{2}\left( Q_{T_{1}}\right) }\leq
C\delta .  \label{7.35}
\end{equation}

By (\ref{7.3})-(\ref{7.5}) and (\ref{7.28})-(\ref{7.34}) 
\begin{equation}
\overline{w}_{t}+a\left( x,t\right) \overline{w}_{xx}=-\widetilde{F}_{t}%
\text{ in }Q_{T_{1}},  \label{7.36}
\end{equation}%
\begin{equation}
\overline{w}\left( 0,t\right) =\overline{w}\left( 1,t\right) =0,\text{ }t\in
\left( 0,T_{1}\right) ,  \label{7.37}
\end{equation}%
\begin{equation}
\overline{w}\left( x,0\right) =0,\text{ }x\in \left( 0,1\right) .
\label{7.38}
\end{equation}%
Also, by (\ref{7.99}), (\ref{7.100}) and (\ref{7.34}) 
\begin{equation}
\widehat{w},\widehat{w}^{\ast },\overline{w}\in H_{0,0}^{2}\left(
Q_{T_{1}}\right) .  \label{7.39}
\end{equation}

Square both sides of equation (\ref{7.36}), multiply by the function $\psi
_{\lambda }^{2}\left( t\right) $ and integrate over the domain $Q_{T_{1}}.$
Using (\ref{7.10}) and (\ref{7.35}), we obtain%
\begin{equation}
\int_{Q_{T_{1}}}\left( \overline{w}_{t}+a\left( x,t\right) \overline{w}%
_{xx}\right) ^{2}\psi _{\lambda }^{2}\left( t\right) dxdt\leq C\delta
^{2}e^{2\left( T_{1}+1\right) ^{\lambda }}.  \label{7.390}
\end{equation}%
Hence, applying Carleman estimate (\ref{7.11}) to the left hand side of (\ref%
{7.390}) and taking into account (\ref{7.10})-(\ref{7.100}), we obtain%
\begin{equation}
\int_{Q_{T_{1}}}\overline{w}_{x}^{2}\psi _{\lambda }^{2}dxdt+\lambda
^{3/2}\int_{Q_{T_{1}}}\overline{w}^{2}\psi _{\lambda }^{2}dxdt\leq
\label{7.41}
\end{equation}%
\begin{equation*}
\leq C\delta ^{2}e^{2\left( T_{1}+1\right) ^{\lambda }}+C\left\Vert 
\overline{w}\right\Vert _{H^{2}\left( Q_{T_{1}}\right) }^{2},\text{ }\forall
\lambda \geq \lambda _{0}.
\end{equation*}%
Since $Q_{T_{1}-\rho }\subset Q_{T_{1}}$ and also since by (\ref{7.9}) $\psi
_{\lambda }^{2}\left( t\right) \geq e^{2\left( T_{1}+1-\rho \right)
^{\lambda }}$ in $Q_{T_{1}-\rho },$ then (\ref{7.41}) implies%
\begin{equation}
\left\Vert \overline{w}_{x}\right\Vert _{L_{2}\left( Q_{T_{1}-\rho }\right)
}^{2}+\left\Vert \overline{w}_{x}\right\Vert _{L_{2}\left( Q_{T_{1}-\rho
}\right) }^{2}\leq  \label{7.42}
\end{equation}%
\begin{equation*}
\leq C\delta e^{\left( T_{1}+1\right) ^{\lambda }}+Ce^{-\left( T_{1}+1-\rho
\right) ^{\lambda }}\left\Vert \overline{w}\right\Vert _{H^{2}\left(
Q_{T_{1}}\right) },\text{ }\forall \lambda \geq \lambda _{0}.
\end{equation*}

Choose $\delta _{0}=\delta _{0}\left( T_{1},a_{0},a_{1}\right) \in \left(
0,1\right) $ so small that 
\begin{equation}
\ln \left[ \ln \left( \delta _{0}^{-1/2}\right) ^{1/\ln \left(
T_{1}+1\right) }\right] >\lambda _{0}.  \label{7.43}
\end{equation}%
Let $\delta \in \left( 0,\delta _{0}\right) .$ We now choose $\lambda
=\lambda \left( \delta \right) $ so large that 
\begin{equation}
e^{\left( T_{1}+1\right) ^{\lambda }}=\frac{1}{\sqrt{\delta }}.
\label{7.430}
\end{equation}%
Hence, 
\begin{equation}
\lambda =\lambda \left( \delta \right) =\ln \left[ \ln \left( \delta
^{-1/2}\right) ^{1/\ln \left( T_{1}+1\right) }\right] >\lambda _{0},\text{ }%
\forall \delta \in \left( 0,\delta _{0}\right) .  \label{7.44}
\end{equation}%
Then 
\begin{equation}
e^{-\left( T_{1}+1-\rho \right) ^{\lambda }}=\exp \left[ -\left( \ln \delta
^{-1/2}\right) ^{\mu }\right] ,  \label{7.45}
\end{equation}%
where the number $\mu \in \left( 0,1\right) $ is defined in (\ref{7.110}).
We have%
\begin{equation}
\frac{e^{-\left( \ln \delta ^{-1/2}\right) ^{\mu }}}{\sqrt{\delta }}=\exp %
\left[ -\frac{1}{2}\ln \delta \left( 1+\frac{2\left( \ln \delta
^{-1/2}\right) ^{\mu }}{\ln \delta }\right) \right] .  \label{1}
\end{equation}%
Since $\mu \in \left( 0,1\right) ,$ then the Hospital's rule implies%
\begin{equation*}
\lim_{\delta \rightarrow 0}\frac{2\left( \ln \delta ^{-1/2}\right) ^{\mu }}{%
\ln \delta }=\lim_{\delta \rightarrow 0}\left( -\mu \left( \ln \delta
^{-1/2}\right) ^{\mu -1}\right) =0.
\end{equation*}%
Hence, 
\begin{equation}
\lim_{\delta \rightarrow 0}\left[ -\frac{1}{2}\ln \delta \left( 1+\frac{%
2\left( \ln \delta ^{-1/2}\right) ^{\mu }}{\ln \delta }\right) \right]
=\lim_{\delta \rightarrow 0}\left( \ln \delta ^{-1/2}\right) =\infty .
\label{2}
\end{equation}%
It follows from (\ref{1}) and (\ref{2}) that%
\begin{equation*}
\lim_{\delta \rightarrow 0}\frac{e^{-\left( \ln \delta ^{-1/2}\right) ^{\mu
}}}{\sqrt{\delta }}=\infty .
\end{equation*}%
Hence, 
\begin{equation}
\sqrt{\delta }\leq C_{1}e^{-\left( \ln \delta ^{-1/2}\right) ^{\mu }},\text{ 
}\forall \delta \in \left( 0,1\right) .  \label{7.450}
\end{equation}

Using (\ref{7.42})-(\ref{7.45}) and (\ref{7.450}), we obtain%
\begin{equation}
\left\Vert \overline{w}_{x}\right\Vert _{L_{2}\left( Q_{T_{1}-\rho }\right)
}+\left\Vert \overline{w}\right\Vert _{L_{2}\left( Q_{T_{1}-\rho }\right)
}\leq  \label{7.46}
\end{equation}%
\begin{equation*}
\leq C_{1}\left( 1+\left\Vert \overline{w}\right\Vert _{H^{2}\left(
Q_{T_{1}}\right) }\right) \exp \left( -\left( \ln \delta ^{-1/2}\right)
^{\mu }\right) ,\text{ }\forall \delta \in \left( 0,\delta _{0}\right) .
\end{equation*}%
By (\ref{7.28})-(\ref{7.32}) $\overline{w}=\left( w-w^{\ast }\right) -%
\widetilde{F}.$ Hence, the triangle inequality, (\ref{7.6}),\emph{\ }(\ref%
{7.28})-(\ref{7.30}), (\ref{7.35}) and (\ref{7.46}) imply (\ref{7.12}),
which is the target estimate of this theorem. $\square $

\subsection{Proof of Theorem 4.3}

\label{sec:4.5}

Denote $\left[ ,\right] $ the scalar product in the space $H^{2}\left(
Q_{T_{1}}\right) .$ Let $\widehat{w}\in H_{0,0}^{2}\left( Q_{T_{1}}\right) $
be the function defined in (\ref{7.31}). Then, using (\ref{7.7}) and (\ref%
{7.340}), consider the functional 
\begin{equation}
I_{\alpha }\left( \widehat{w}+F\right) =\int_{Q_{T_{1}}}\left( P\widehat{w}%
+F_{t}\right) ^{2}dxdt+\alpha \left\Vert \widehat{w}+F\right\Vert
_{H^{2}\left( Q_{T_{1}}\right) }^{2}.  \label{7.47}
\end{equation}%
Suppose that the function $\widehat{w}_{\min }\in H_{0,0}^{2}\left(
Q_{T_{1}}\right) $ is a minimizer of the functional $I_{\alpha }\left( 
\widehat{w}+F\right) $ on the space $H_{0,0}^{2}\left( Q_{T_{1}}\right) ,$
i.e.%
\begin{equation}
I_{\alpha }\left( \widehat{w}_{\min }+F\right) \leq I_{\alpha }\left( 
\widehat{w}+F\right) ,\text{ }\forall \widehat{w}\in H_{0,0}^{2}\left(
Q_{T_{1}}\right) .  \label{7.48}
\end{equation}%
Consider the function $w_{\min }=\widehat{w}_{\min }+F.$ Then it follows
from (\ref{7.100}) and (\ref{7.29}) that $w_{\min }\in Y,$ where the set $Y$
is defined in (\ref{7.8}). Also, $w=\widehat{w}+F\in Y,$ $\forall \widehat{w}%
\in H_{0,0}^{2}\left( Q_{T_{1}}\right) .$ Hence, (\ref{7.48}) implies that
the function $w_{\min }$ is a minimizer of the functional $I_{\alpha }\left(
w\right) $ on the set $Y$.

We now prove the reverse. Suppose that a function $w^{\min }\in Y$ is a
minimizer of the functional $I_{\alpha }\left( w\right) $ on the set $Y$,
i.e.%
\begin{equation}
I_{\alpha }\left( w^{\min }\right) \leq I_{\alpha }\left( w\right) ,\text{ }%
\forall w\in Y.  \label{7.49}
\end{equation}%
Consider the function $\widehat{w}^{\min }=w^{\min }-F.$ And for every
function $\widehat{w}\in H_{0,0}^{2}\left( Q_{T_{1}}\right) $ consider the
function $w=\widehat{w}+F\in Y.$ Then by (\ref{7.49})%
\begin{equation*}
I_{\alpha }\left( \widehat{w}^{\min }+F\right) =I_{\alpha }\left( w^{\min
}\right) \leq I_{\alpha }\left( w\right) =I_{\alpha }\left( \widehat{w}%
+F\right) ,\text{ }\forall \widehat{w}\in H_{0,0}^{2}\left( Q_{T_{1}}\right)
.
\end{equation*}%
Hence, $\widehat{w}^{\min }$ is a minimizer of functional (\ref{7.47}) on
the space $H_{0,0}^{2}\left( Q_{T_{1}}\right) .$ Therefore, it is sufficient
to find a minimizer of the functional $I_{\alpha }\left( \widehat{w}%
+F\right) $ on the space $H_{0,0}^{2}\left( Q_{T_{1}}\right) .$

By the variational principle the function $\widehat{w}_{\min }\in
H_{0,0}^{2}\left( Q_{T_{1}}\right) $ is a minimizer of the functional $%
I_{\alpha }\left( \widehat{w}+F\right) $ if and only if the following
integral identity is satisfied:%
\begin{equation}
\int_{Q_{T_{1}}}\left( P\widehat{w}_{\min }\cdot Ph\right) dxdt+\alpha \left[
\widehat{w}_{\min },h\right] =-\int_{Q_{T_{1}}}F_{t}\cdot Phdxdt-\alpha %
\left[ F,h\right] ,\text{ }  \label{7.50}
\end{equation}%
\begin{equation*}
\forall h\in H_{0,0}^{2}\left( Q_{T_{1}}\right) .
\end{equation*}%
Consider a new scalar product in $H_{0,0}^{2}\left( Q_{T_{1}}\right) $
defined as 
\begin{equation*}
\left\{ u,v\right\} =\int_{Q_{T_{1}}}\left( Pu\cdot Pv\right) dxdt+\alpha 
\left[ u,v\right] ,\text{ }\forall u,v\in H_{0,0}^{2}\left( Q_{T_{1}}\right)
.
\end{equation*}%
Recall that $\alpha \in \left( 0,1\right) .$ Obviously, 
\begin{equation*}
\alpha \left\Vert u\right\Vert _{H^{2}\left( Q_{T_{1}}\right) }^{2}\leq
\left\{ u,u\right\} \leq C\left\Vert u\right\Vert _{H^{2}\left(
Q_{T_{1}}\right) }^{2},\forall u\in H_{0,0}^{2}\left( Q_{T_{1}}\right) .
\end{equation*}%
Hence, norms $\sqrt{\left\{ u,u\right\} }$ and $\left\Vert u\right\Vert
_{H^{2}\left( Q_{T_{1}}\right) }$ are equivalent. Hence, one can consider
the scalar product $\left\{ u,v\right\} $ as the scalar product in $%
H_{0,0}^{2}\left( Q_{T_{1}}\right) .$

Hence, we can rewrite (\ref{7.50}) as 
\begin{equation}
\left\{ \widehat{w}_{\min },h\right\} =-\int_{Q_{T_{1}}}F_{t}\cdot
Phdxdt-\alpha \left[ F,h\right] ,\text{ }\forall h\in H_{0,0}^{2}\left(
Q_{T_{1}}\right) .  \label{7.51}
\end{equation}%
The right hand side of (\ref{7.51}) can be estimated as%
\begin{equation*}
\left\vert -\int_{Q_{T_{1}}}F_{t}Phdxdt-\alpha \left[ F,h\right] \right\vert
\leq C\left\Vert F\right\Vert _{H^{2}\left( Q_{T_{1}}\right) }\left\{
h\right\} ,\text{ }\forall h\in H_{0,0}^{2}\left( Q_{T_{1}}\right) .
\end{equation*}%
Hence, the right hand side of (\ref{7.51}) can be considered as a bounded
linear functional of $h\in H_{0,0}^{2}\left( Q_{T_{1}}\right) .$ Hence, by
Riesz theorem there exists unique function $W\in H_{0,0}^{2}\left(
Q_{T_{1}}\right) $ such that 
\begin{equation*}
-\int_{Q_{T_{1}}}F_{t}Phdxdt-\alpha \left[ F,h\right] =\left\{ W,h\right\} ,%
\text{ }\forall h\in H_{0,0}^{2}\left( Q_{T_{1}}\right) .
\end{equation*}%
Comparing this with (\ref{7.51}), we obtain 
\begin{equation*}
\left\{ \widehat{w}_{\min },h\right\} =\left\{ W,h\right\} ,\text{ }\forall
h\in H_{0,0}^{2}\left( Q_{T_{1}}\right) .
\end{equation*}%
Therefore, $\widehat{w}_{\min }=W.$ Thus, we have proven existence and
uniqueness of the minimizer $\widehat{w}_{\min }$ of the functional $%
I_{\alpha }\left( \widehat{w}+F\right) $ on the space $H_{0,0}^{2}\left(
Q_{T_{1}}\right) .$ Therefore, it follows from the discussion in the
beginning of this proof that there exists unique minimizer \ of the
functional $I_{\alpha }\left( w\right) $ on the set $Y$ and this minimizer
is $w_{\min }=\widehat{w}_{\min }+F.$

We now estimate the norm $\left\Vert w_{\min }\right\Vert _{H^{2}\left(
Q_{T_{1}}\right) }.$ Setting in (\ref{7.50}) $h=\widehat{w}_{\min }$ and
using Cauchy-Schwarz inequality, we obtain 
\begin{equation*}
\int_{Q_{T_{1}}}\left( P\widehat{w}_{\min }\right) ^{2}dxdt+\alpha
\left\Vert \widehat{w}_{\min }\right\Vert _{H^{2}\left( Q_{T_{1}}\right)
}^{2}\leq
\end{equation*}%
\begin{equation*}
\leq \frac{1}{2}\left\Vert F_{t}\right\Vert _{L_{2}\left( Q_{T_{1}}\right)
}^{2}+\frac{1}{2}\left\Vert P\widehat{w}_{\min }\right\Vert _{H^{2}\left(
Q_{T_{1}}\right) }^{2}+\frac{\alpha }{2}\left\Vert F\right\Vert
_{H^{2}\left( Q_{T_{1}}\right) }^{2}+\frac{\alpha }{2}\left\Vert \widehat{w}%
_{\min }\right\Vert _{H^{2}\left( Q_{T_{1}}\right) }^{2}.
\end{equation*}%
Hence,%
\begin{equation*}
\left\Vert \widehat{w}_{\min }\right\Vert _{H^{2}\left( Q_{T_{1}}\right)
}\leq \frac{C}{\sqrt{\alpha }}\left\Vert F\right\Vert _{H^{2}\left(
Q_{T_{1}}\right) }.
\end{equation*}%
This estimate, triangle inequality and (\ref{7.29}) imply the target
estimate (\ref{7.13}) of Theorem 4.3. $\square $

\subsection{Proof of Theorem 4.4}

\label{sec:4.6}

We still use notations (\ref{7.28})-(\ref{7.32}). By Corollary 4.1 Problem 3
has at most one solution. Hence, there exists unique exact solution $w^{\ast
}\in H^{2}\left( Q_{T_{1}}\right) $ of Problem 3 with the data $\varphi
_{0}^{\ast },\varphi _{1}^{\ast }\in H^{2}\left( 0,T_{1}\right) $ in (\ref%
{7.4}) and (\ref{7.5}). Hence, we have the following analog of integral
identity (\ref{7.50}) 
\begin{equation}
\int_{Q_{T_{1}}}\left( P\widehat{w}^{\ast }\cdot Ph\right) dxdt+\alpha \left[
\widehat{w}^{\ast },h\right] =  \label{7.53}
\end{equation}%
\begin{equation*}
=-\int_{Q_{T_{1}}}F_{t}^{\ast }\cdot Phdxdt+\alpha \left[ \widehat{w}^{\ast
},h\right] ,\text{ }\forall h\in H_{0,0}^{2}\left( Q_{T_{1}}\right) .
\end{equation*}%
Subtract (\ref{7.53}) from (\ref{7.50}). Then, using (\ref{7.30}), (\ref%
{7.32}) and (\ref{7.52}), we obtain%
\begin{equation}
\int_{Q_{T_{1}}}\left( P\overline{w}\cdot Ph\right) dxdt+\alpha \left[ 
\overline{w},h\right] =  \label{7.54}
\end{equation}%
\begin{equation*}
=-\int_{Q_{T_{1}}}\widetilde{F}_{t}\cdot Phdxdt+\alpha \left[ \widehat{w}%
^{\ast },h\right] ,\text{ }\forall h\in H_{0,0}^{2}\left( Q_{T_{1}}\right) .
\end{equation*}%
Set in (\ref{7.54}) $h=\overline{w}.$ Then, using (\ref{7.35}) and
Cauchy-Schwarz inequality, we obtain%
\begin{equation}
\int_{Q_{T_{1}}}\left( P\overline{w}\right) ^{2}dxdt\leq C\left( \delta
^{2}+\alpha \left\Vert \widehat{w}^{\ast }\right\Vert _{H^{2}\left(
Q_{T_{1}}\right) }^{2}\right) ,  \label{7.55}
\end{equation}%
\begin{equation}
\left\Vert \overline{w}\right\Vert _{H^{2}\left( Q_{T_{1}}\right) }\leq
C\left( \frac{\delta }{\sqrt{\alpha }}+\left\Vert \widehat{w}^{\ast
}\right\Vert _{H^{2}\left( Q_{T_{1}}\right) }\right) .  \label{7.56}
\end{equation}%
Inequality (\ref{7.55}) is equivalent with%
\begin{equation*}
\int_{Q_{T_{1}}}\left( P\overline{w}\right) ^{2}\psi _{\lambda }^{2}\psi
_{\lambda }^{-2}dxdt\leq C\left( \delta ^{2}+\alpha \left\Vert \widehat{w}%
^{\ast }\right\Vert _{H^{2}\left( Q_{T_{1}}\right) }^{2}\right) .
\end{equation*}%
Since by (\ref{7.10}) $\psi _{\lambda }^{-2}\left( t\right) \geq e^{-2\left(
T_{1}+1\right) ^{\lambda }}$ in $Q_{T_{1}},$ then (\ref{7.55}) implies%
\begin{equation}
\int_{Q_{T_{1}}}\left( P\overline{w}\right) ^{2}\psi _{\lambda }^{2}dxdt\leq
C\left( \delta ^{2}+\alpha \left\Vert \widehat{w}^{\ast }\right\Vert
_{H^{2}\left( Q_{T_{1}}\right) }^{2}\right) e^{2\left( T_{1}+1\right)
^{\lambda }}.  \label{7.57}
\end{equation}%
Hence, applying Carleman estimate (\ref{7.11}) to the left hand side of (\ref%
{7.57}) and recalling again that $\alpha \in \left( 0,1\right) $, we obtain 
\begin{equation*}
\int_{Q_{T_{1}}}\overline{w}_{x}^{2}\psi _{\lambda }^{2}dxdt+\lambda
^{3/2}\int_{Q_{T_{1}}}\overline{w}^{2}\psi _{\lambda }^{2}dxdt\leq
\end{equation*}%
\begin{equation*}
\leq C\left( \delta ^{2}+\alpha \left\Vert \widehat{w}^{\ast }\right\Vert
_{H^{2}\left( Q_{T_{1}}\right) }^{2}\right) e^{2\left( T_{1}+1\right)
^{\lambda }}+C\left\Vert \overline{w}\right\Vert _{H^{2}\left(
Q_{T_{1}}\right) }^{2},\text{ }\forall \lambda \geq \lambda _{0}.
\end{equation*}%
Hence, we obtain similarly with (\ref{7.42})%
\begin{equation*}
\left\Vert \overline{w}_{x}\right\Vert _{L_{2}\left( Q_{T_{1}-\rho }\right)
}^{2}+\left\Vert \overline{w}\right\Vert _{L_{2}\left( Q_{T_{1}-\rho
}\right) }^{2}\leq C\left( \delta ^{2}+\alpha \left\Vert \widehat{w}^{\ast
}\right\Vert _{H^{2}\left( Q_{T_{1}}\right) }^{2}\right) e^{2\left(
T_{1}+1\right) ^{\lambda }}
\end{equation*}%
\begin{equation*}
+Ce^{-2\left( T_{1}+1-\rho \right) ^{\lambda }}\left\Vert \overline{w}%
\right\Vert _{H^{2}\left( Q_{T_{1}}\right) }^{2},\text{ }\forall \lambda
\geq \lambda _{0}.
\end{equation*}%
Combining this with (\ref{7.56}), we obtain 
\begin{equation}
\left\Vert \overline{w}_{x}\right\Vert _{L_{2}\left( Q_{T_{1}-\rho }\right)
}+\left\Vert \overline{w}\right\Vert _{L_{2}\left( Q_{T_{1}-\rho }\right)
}\leq C\left( \delta +\sqrt{\alpha }\left\Vert \widehat{w}^{\ast
}\right\Vert _{H^{2}\left( Q_{T_{1}}\right) }\right) e^{\left(
T_{1}+1\right) ^{\lambda }}  \label{7.58}
\end{equation}%
\begin{equation*}
+C\frac{\delta }{\sqrt{\alpha }}e^{-\left( T_{1}+1-\rho \right) ^{\lambda
}}+\left\Vert \widehat{w}^{\ast }\right\Vert _{H^{2}\left( Q_{T_{1}}\right)
}e^{-\left( T_{1}+1-\rho \right) ^{\lambda }},\text{ }\forall \lambda \geq
\lambda _{0}.
\end{equation*}%
Suppose now that $\alpha =\alpha \left( \delta \right) =\delta ^{2},$ as
stated in (\ref{7.140}). Choose $\delta _{0}=\delta _{0}\left(
T_{1},a_{0},a_{1}\right) \in \left( 0,1\right) $ as in (\ref{7.43}) and $%
\lambda =\lambda \left( \delta \right) $ as in (\ref{7.44}). Then (\ref%
{7.430}), (\ref{7.45}), (\ref{7.450}) and (\ref{7.58}) imply 
\begin{equation}
\left\Vert \overline{w}_{x}\right\Vert _{L_{2}\left( Q_{T_{1}-\rho }\right)
}+\left\Vert \overline{w}\right\Vert _{L_{2}\left( Q_{T_{1}-\rho }\right)
}\leq  \label{7.64}
\end{equation}%
\begin{equation*}
\leq C_{1}\left( 1+\left\Vert \widehat{w}^{\ast }\right\Vert _{H^{2}\left(
Q_{T_{1}}\right) }\right) \exp \left[ -\left( \ln \delta ^{-1/2}\right)
^{\mu }\right] ,\text{ }\forall \delta \in \left( 0,\delta _{0}\right) .
\end{equation*}%
The target estimate (\ref{7.14}) of this theorem follows immediately from
the triangle inequality, (\ref{7.6}), (\ref{7.28})-(\ref{7.32}) and (\ref%
{7.64}). $\square $

\section{Probabilistic Arguments for a Trading Strategy}

\label{sec:5}

A heuristic algorithm of section 3 can be used as the basis for a trading
strategy of options. The algorithm predicts the option price change
relatively to the current price. The fact that this algorithm uses the
information about stock and option prices only over a small time period
makes realistic the assumptions of the model of Section 2 about the
volatilities being independent on time. Formulas (\ref{2.4}) and (\ref{2.7})
indicate that the sign of the mathematical expectation of the option price
increment should likely define the trading strategy. In addition to the
mathematical expectation of the option price increment, it is necessary to
take into account indicators that reflect the risk of using that trading
strategy. This is because the option price dynamics is described by a random
process. Based on the model of Section 2, we construct in this section such
indicators for a "perfect" trading strategy, which always correctly predicts
the sign of the mathematical expectation of the option price increment.

We assume in this section that both the volatility $\sigma $ of the stock
and the idea $\hat{\sigma}$ of the volatility of the call option, which has
been developed among the participants involved in trading of this option,
are known. Recall that we have assumed in Section 2 that the dynamics of the
stock price is described by a stochastic differential equation of the
geometric Brownian motion $ds=\sigma sdW$ with the initial condition $%
s(t_{0})=s_{0}$, and also that the corresponding option price is $%
v(s(t),t)=u(s(t),T-\tau )$ , where $\tau =T-t\in \left( 0,T\right) $ and the
function $u(s,\tau )$ can be found by the Black-Scholes formula (\ref{2.3}).
The option price expected by option market participants is described by a
stochastic process $v(\hat{s}(t),t)=u(\hat{s}(t),T-t)$, where the expected
stock price satisfies the stochastic differential equation of the geometric
Brownian motion $d\hat{s}=\hat{\sigma}\hat{s}d\hat{W}_{1}$ with the initial
condition 
\begin{equation}
\hat{s}(t_{0})=s_{0}=s(t_{0}).  \label{4.00}
\end{equation}%
Here $W_{1}$ is a Wiener process, and the processes $W$ and $W_{1}$ are
independent.

Let $t_{0}\in \left( 0,T\right) $ be a certain moment of time and $%
\varepsilon >0$ be a sufficiently small number. The true option price at the
moment of time $t_{0}+\varepsilon $ is $v(s(t_{0}+\varepsilon
),t_{0}+\varepsilon ).$ On the other hand, at the same moment of time $%
t_{0}+\varepsilon $ the price of this option expected by the participants of
the market is $v(\hat{s}(t_{0}+\varepsilon ),t_{0}+\varepsilon )$. It
follows from (\ref{2.4}) that, on the small time interval $%
(t_{0},t_{0}+\varepsilon ),$ a winning trading strategy of the options
trading should be based on an estimate of the probability that $%
v(s(t_{0}+\varepsilon ),t_{0}+\varepsilon )>v(\hat{s}(t_{0}),t_{0})$. This
probability is given in Theorem 5.1.

\textbf{Theorem 5.1}. \emph{Let }$\varepsilon >0$\emph{\ be a sufficiently
small number and }$\Phi \left( z\right) ,z\in \mathbb{R}$\emph{\ be the
function defined in (\ref{2.200}). The probability that at the time }$%
t_{0}+\varepsilon $\emph{\ the true option price }$v(s(t_{0}+\varepsilon
),t_{0}+\varepsilon )$\emph{\ is greater than the price expected by the
participants of the options market }$v(\hat{s}(t_{0}+\varepsilon
),t_{0}+\varepsilon )$\emph{\ is }%
\begin{equation}
p=\Phi \left( \frac{(\hat{\sigma}^{2}-\sigma ^{2})\sqrt{\varepsilon }}{2%
\sqrt{(\hat{\sigma}^{2}+\sigma ^{2})}}\right) .  \label{4.1}
\end{equation}

\textbf{Proof.} The derivative 
\begin{equation*}
\frac{\partial u(s,\tau )}{\partial s}
\end{equation*}%
is called the Greek delta. This parameter for a call option is 
\begin{equation}
\Delta =\frac{\partial u(s,\tau )}{\partial s}=\Phi (\Theta _{+}(s,\tau ))>0.
\label{4.2}
\end{equation}%
Since by (\ref{4.2}) $\partial _{s}u(s,\tau )>0,$ then the inequality $%
v(s(t_{0}+\varepsilon ),t_{0}+\varepsilon )>v(\hat{s}(t_{0}+\varepsilon
),t_{0}+\varepsilon )$ is equivalent to the inequality $s(t_{0}+\varepsilon
)>\hat{s}(t_{0}+\varepsilon )$. It follows from (\ref{4.00}) that the latter
inequality is equivalent with 
\begin{equation*}
\ln \left( {\frac{s(t_{0}+\varepsilon )}{s(t_{0})}}\right) >\ln \left( {%
\frac{\hat{s}(t_{0}+\varepsilon )}{\hat{s}(t_{0})}}\right) .
\end{equation*}

It follows from the properties of the geometric Brownian motion, see, e.g. 
\cite[Chapter 5, section 5.1]{Bernt} that the random variables 
\begin{equation*}
\ln \left( \frac{s(t_{0}+\varepsilon )}{s(t_{0})}\right) \in N\left( -\frac{%
\sigma ^{2}}{2}\varepsilon ,\sigma ^{2}{\varepsilon }\right) ,
\end{equation*}

\begin{equation*}
\ln \left( \frac{\hat{s}(t_{0}+\varepsilon )}{\hat{s}(t_{0})}\right) \in
N\left( -\frac{\hat{\sigma}^{2}}{2}\varepsilon ,\hat{\sigma}^{2}{\varepsilon 
}\right)
\end{equation*}%
are normally distributed. Hence, the difference of these two random
variables is also a normally distributed random variable, see, e.g. \cite[%
Chapter 9, section 9.3]{KorSin}, i.e.

\begin{equation*}
\left[ \ln \left( \frac{s(t_{0}+\varepsilon )}{s(t_{0})}\right) -\ln \left( 
\frac{\hat{s}(t_{0}+\varepsilon )}{\hat{s}(t_{0})}\right) \right] \in
N\left( \frac{\sigma ^{2}-\hat{\sigma}^{2}}{2}\varepsilon ,(\hat{\sigma}%
^{2}+\sigma ^{2})\varepsilon \right) .
\end{equation*}

Therefore, the value given by formula (\ref{4.1}) is indeed the probability
that the true option price $v(s(t_{0}+\varepsilon ),t_{0}+\varepsilon ))$ is
greater than the expected option price $v(\hat{s}(t_{0}+\varepsilon
),t_{0}+\varepsilon ).$ $\square $

Theorem 5.1 implies that the operation of buying an option at the time
moment $t_{0}$ and selling it at the time moment $t_{0}+\varepsilon $ will
be profitable with the probability $p$ given in (\ref{4.1}), and this
operation will be non-profitable with the probability $1-p.$

By (\ref{2.200}) and (\ref{4.1})%
\begin{equation*}
p=\left\{ 
\begin{array}{c}
>1/2\text{ if }\sigma >\hat{\sigma}, \\ 
\leq 1/2\text{ if }\sigma \leq \hat{\sigma}.%
\end{array}%
\right.
\end{equation*}%
Hence, if $\sigma >\hat{\sigma}$, then it is reasonable to buy an option at
the moment of time $t_{0}$ and sell it at the moment of time $%
t_{0}+\varepsilon .$ Otherwise, it is reasonable to go in the short position
on the option at $t=t_{0}$. Suppose that $p>1/2$. A winning strategy, which
takes into account risks, should involve the repetition operation multiple
times with the same independent probabilities of outcomes. Consider $n$
non-overlapping small time intervals $[t_{j},t_{j}+\varepsilon ],$ $%
j=1,...,n,$ of the same duration $\varepsilon >0,$ on which the option
purchase operations are carried out at the moment of time $t_{j}$ with the
subsequent sale at the moment of time $t_{j}+\varepsilon $.

Consider random variables $\left\{ \xi _{j}\right\} _{j=1}^{n},$

\begin{equation}
\xi _{j}=%
\begin{cases}
1, & \text{if }{v(s(t_{j}+\varepsilon ),t_{j}+\varepsilon )\geq v(\hat{s}%
(t_{j}),t_{j}),} \\ 
0, & \text{if }{v(s({t_{j}+}\varepsilon ),{t_{j}+}\varepsilon )<v(\hat{s}({%
t_{j}}),{t_{j}})},%
\end{cases}%
j=1,...,n.  \label{3.99}
\end{equation}

If $\xi _{j}=1,$ then the operation of buying that option at the moment of
time $t_{j}$ and selling it at the moment of time $t_{j}+\varepsilon $ was
profitable. If $\xi _{j}=0,$ then that operation was non profitable. The
random variables $\xi _{1},...,\xi _{n}$ are independent identically
distributed random variables \cite[Chapter 18, section 18.1]{KorSin}. The
frequency of profitable trading operations is characterized by the random
variable $\zeta ,$

\begin{equation}
\zeta =\frac{1}{n}\sum_{j=1}^{n}\xi _{j}.  \label{4.0}
\end{equation}%
It follows from \cite[Chapter 9, section 9.3]{KorSin} that the variable $%
\zeta $ has a binomial distribution with the mathematical expectation $p$
given in (\ref{4.1}) and with the dispersion $D$, where 
\begin{equation}
D=\frac{p(1-p)}{n}.  \label{4.01}
\end{equation}%
By the Central Limit Theorem of de Moivre-Laplace, the probability that more
than half of trades are profitable is estimated as \cite[Chapter 2, section
2.2]{KorSin}:

\begin{equation}
\sum_{k=[\frac{n}{2}+2]}^{n}\frac{n!}{(k-1)!(n-k)!}p^{k}(1-p)^{n-k}=\Phi
\left( \frac{(1-2p)\sqrt{n}}{2\sqrt{p(1-p)}}\right) +\delta (n),  \label{4.3}
\end{equation}

\begin{equation}
\lim_{n\rightarrow \infty }\delta (n)=0.  \label{4.4}
\end{equation}%
In our trading strategy, we decide to make transactions if and only if the
probability of the profitable trading is not less than a given value $\alpha
>1/2.$ Hence, by (\ref{4.3}) and (\ref{4.4})%
\begin{equation}
\frac{(1-2p)\sqrt{n}}{2\sqrt{p(1-p)}}>\Phi ^{-1}\left( \alpha -\delta
(n)\right) ,  \label{4.40}
\end{equation}%
where $\Phi ^{-1}$ is the inverse function of the function $\Phi $ of (\ref%
{2.200}). By (\ref{4.40}) 
\begin{equation*}
p^{2}-p+\frac{n}{4\left\{ n+\left[ \Phi ^{-1}(\alpha -\delta (n)\right]
^{2}\right\} }>0.
\end{equation*}%
Thus, we should have

\begin{equation}
p\geq \frac{1}{2}\left( 1+\sqrt{\frac{\left[ \Phi ^{-1}(\alpha -\delta (n)%
\right] ^{2}}{n+\left[ \Phi ^{-1}(\alpha -\delta (n)\right] ^{2}}}\right) .
\label{4.5}
\end{equation}

To fulfill inequality (\ref{4.5}), the imperfection of the stock market must
be significant. More precisely, it follows from (\ref{4.1}) and (\ref{4.5})
that the difference between the volatilities $\sigma $ and $\hat{\sigma}$
must satisfy the following inequality:

\begin{equation}
\sigma -\hat{\sigma}\geq \frac{2\sqrt{\sigma ^{2}+\hat{\sigma}^{2}}}{\sqrt{%
\varepsilon }(\sigma +\hat{\sigma})}\left\vert \Phi ^{-1}\left( \frac{1}{2}%
\left( 1+\sqrt{\frac{\left[ \Phi ^{-1}(\alpha -\delta (n)\right] ^{2}}{n+%
\left[ \Phi ^{-1}(\alpha -\delta (n)\right] ^{2}}}\right) \right)
\right\vert .  \label{4.6}
\end{equation}

Based on this estimate of the difference $\sigma -\hat{\sigma}$, we design a
trading strategy in the ideal case. \textquotedblleft Ideal" means that we
know both volatilities $\sigma $ and $\hat{\sigma}.$

\textbf{Trading Strategy for the Ideal Case:}

Let $\beta _{1}>0$ and $\beta _{2}<0$ be two threshold numbers. Our trading
strategy considers three possible scenarios:

\begin{enumerate}
\item If $\sigma -\hat{\sigma}\geq \beta _{1}>0$ , then it is recommended to
buy a call option at the current moment of time $t_{j}$ with the subsequent
sale at the next moment of time $t_{j}+\varepsilon $.

\item If $\sigma -\hat{\sigma}\leq \beta _{2}$, then then it is recommended
to go short at the current moment of time $t_{j}$, followed by closing the
short position at time $t_{j}+\varepsilon $.

\item If $\beta _{2}<\sigma -\hat{\sigma}<\beta _{1}$, then it is
recommended to refrain from trading.
\end{enumerate}

The threshold values $\beta _{1}$ and $\beta _{2}$ might probably estimated
via numerical simulations using the method of section 3, combined with
formula (\ref{4.6}).

\section{Numerical Studies}

\label{sec:6}

\subsection{Some numerical details for the algorithm of section 3}

\label{sec:6.1}

We have computationally simulated the market data as described in subsection
6.2. These data gave us initial and boundary conditions (\ref{3.1}), (\ref%
{3.3}) and (\ref{3.4}), which, in turn, led us to (\ref{3.11}), (\ref{3.12}%
), see Steps 1,2 of subsection 3.2. Next, we have solved Minimization
Problem (\ref{3.14}), (\ref{3.15}). To minimize functional (\ref{3.14}), we
wrote $Mv$ and $\left\Vert v\right\Vert _{H^{2}\left( G_{2y}\right) }^{2}$
in finite differences and, using the conjugate gradient method, have
minimized the resulting discrete functional with respect to the values of
the function $v$ at grid points. The starting point of the minimization
procedure was $v=0$. The regularization parameter $\gamma =0.01$ was the
same as in \cite{KlibGol}, and it was chosen by trial and error. The step
sizes $h_{x}$ and $h_{t}$ of the finite difference scheme with respect to $x$
and $t$ were $h_{x}=0.01$ and $h_{t}=0.0000784$ respectively. Since by (\ref%
{3.100}) and (\ref{3.8}) $G_{2y}=\left\{ \left( x,t\right) \in \left(
0,1\right) \times \left( 0,0.00784\right) \right\} ,$ then we had 100 grid
points with respect to each variable $x$ and $t.$

\subsection{The data}

\label{sec:6.2}

We construct the stock price trajectory $s(t)$ as a solution to the
stochastic differential equation $ds=\sigma sdW$ with the initial condition $%
s(0)=100,$ where $\sigma =0.2.$ We model the stock prices and then the
prices of 90-days European call options of this stock during the life of the
stock, assuming that the options are reissued many times with the same
maturity date of 90 days. Thus, we obtain a time series $\left\{ s\left(
t_{k}\right) \right\} _{k=1}^{N},N\geq n=2000.$ We set the payoff function
for each option $f(s)=(s-100)_{+}.$ The generated stock price trajectory is
shown on Figure ~\ref{fig: An example of the time dependent behavior of
stock prices generated by the geometric Brownian motion}.

\begin{figure}[h]
\par
\begin{center}
{\includegraphics[width = 1.1\textwidth, height =
0.60\textwidth]{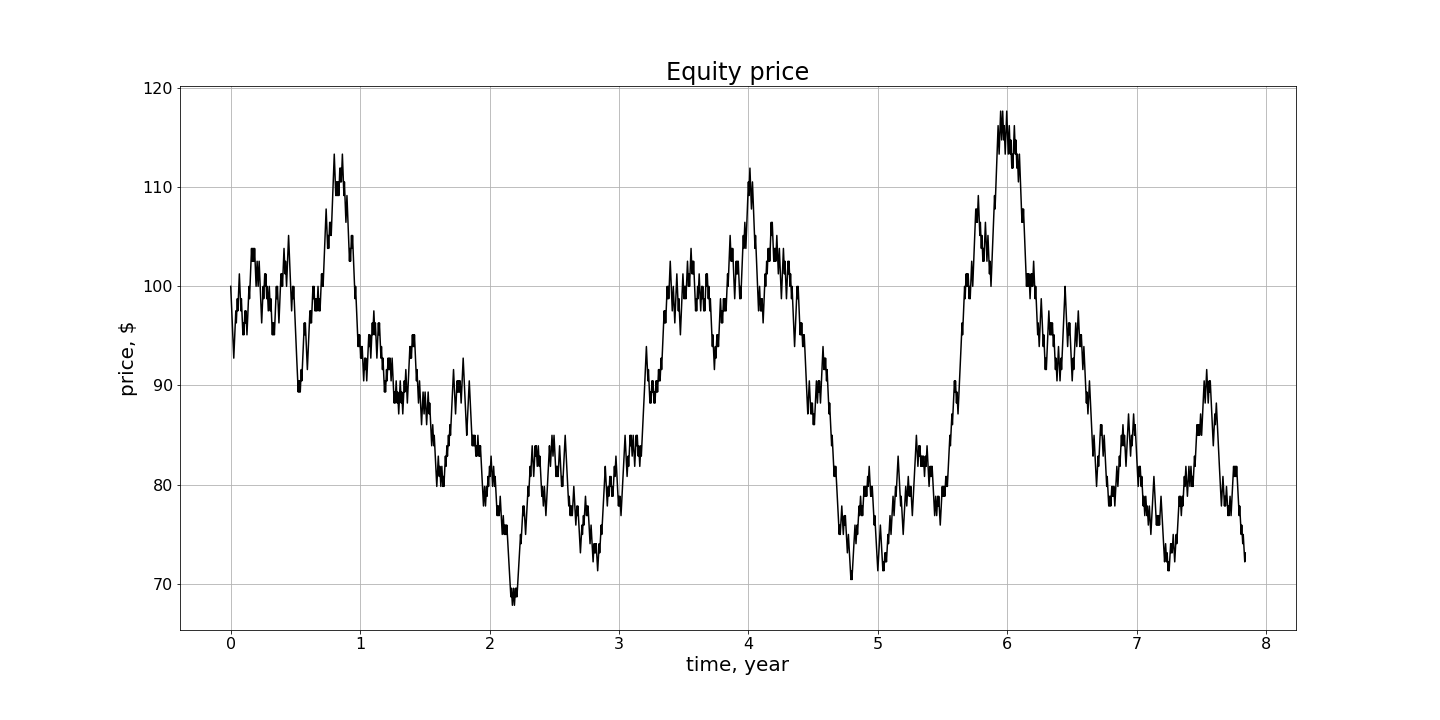}}
\par
\end{center}
\caption{\emph{\ An example of the time dependent behavior of stock prices
generated by the geometric Brownian motion with $\protect\sigma =0.2.$ }}
\label{fig:  An example of the time dependent behavior of stock prices
generated by the geometric Brownian motion}
\end{figure}

The probabilistic analysis of section 5 of the random variable $\zeta $
characterizes the effectiveness of an \textquotedblleft ideal" trading
strategy. Thus, we consider the ideal case first. Recall that
\textquotedblleft ideal" means here that this strategy is based on the
knowledge of the information about the imperfection of the financial market,
i.e. on the knowledge of both volatilities $\sigma $ and $\hat{\sigma}$.
However, in the real market data only approximate values of\textbf{\ }$\hat{%
\sigma}$ are available \cite{Bloom}.

We test total thirty three (33) values of $\hat{\sigma}.$ More precisely, in
our computational simulations, we took the discrete values of $\hat{\sigma},$
where 
\begin{equation}
\hat{\sigma}\in \lbrack 0.05,0.38]\text{ with the step size }h_{\hat{\sigma}%
}=0.01.  \label{6.1}
\end{equation}%
We now generate the function, which describes the dependence of the
mathematical expectation of the random variable $\zeta $ in (\ref{4.0}) on $%
\hat{\sigma}$. Keeping in mind that $\sigma =0.2$ in all cases, we compute
for each of the discrete values of $\hat{\sigma}$ in (\ref{6.1}) the
mathematical expectation $p\left( \hat{\sigma}\right) $ of the random
variable $\zeta $. We use formula (\ref{4.1}) for $p\left( \hat{\sigma}%
\right) ,$ also, see (\ref{2.200}). This way we obtain the function $p\left( 
\hat{\sigma}\right) ,$ which is the above dependence for the ideal case.

Second, we consider a non-ideal case. More precisely, \ we test how our
heuristic algorithm works for the computationally simulated data described
in this subsection. We choose $n=2000$ non-overlapping time intervals $%
[t_{j},t_{j}+y],$ $j=1,...,n,$ where $y=1/255$ means one dimensionless
trading day, see (\ref{1.2}) and (\ref{3.100}). We still use the
dimensionless time as in (\ref{1.1}), while keeping the same notation $t$
for brevity. For every fixed value of $\hat{\sigma}$ indicated in (\ref{6.1}%
), we calculate the option price $v(s\left( t_{j}\right) ,t_{j})=u(s\left(
T-t_{j}\right) ,T-t_{j}),$ where the function $u\left( s,\tau \right)
=u\left( s,T-t\right) $ is given by Black-Scholes formula (\ref{2.2}). Thus,
numbers $v(s\left( t_{j}\right) ,t_{j})$ form the option price trajectory.
Figure 2 displays a sample of the trajectory of the option price for $\hat{%
\sigma}=0.1.$ Based on (\ref{3.0}), we set bid and ask stock prices as well
as corresponding bid and ask option prices as: 
\begin{equation*}
s_{b}\left( t_{j}\right) =0.99\cdot s\left( t_{j}\right) \text{ and }%
s_{a}\left( t_{j}\right) =1.01\cdot s\left( t_{j}\right) ,
\end{equation*}
\begin{equation*}
v_{b}(s\left( t_{j}\right) ,t_{j})=0.99\cdot v(s\left( t_{j}\right) ,t_{j})%
\text{ and }v_{a}(s\left( t_{j}\right) ,t_{j})=1.01\cdot v(s\left(
t_{j}\right) ,t_{j}).
\end{equation*}%
Next, we solve Problem 2 of section 3 for each $j$ on the time interval $%
\left[ t_{j},t_{j}+2y\right] $ by the algorithm of that section. When doing
so, we take in (\ref{3.6}) $\sigma ^{2}\left( t\right) =\hat{\sigma}^{2}$
for $t\in \left[ t_{j},t_{j}+2y\right] $ for all $j=1,...,2000.$ In
particular, this solution via QRM gives us the function $v_{\text{comp}%
}\left( s,t_{j}+y\right) ,$ $s\in \left( s_{b}\left( t_{j}\right)
,s_{a}\left( t_{j}\right) \right) .$ We set the predicted price of the
option at the moment of time $t_{j}+y$ as: 
\begin{equation}
v_{\text{pred}}\left( t_{j}+y\right) =v_{\text{comp}}\left( \frac{%
s_{b}\left( t_{j}\right) +s_{a}\left( t_{j}\right) }{2},t_{j}+y\right) .
\label{6.2}
\end{equation}%
\clearpage
\begin{figure}[h]
\par
\begin{center}
{\includegraphics[width = 1.1\textwidth, height =
0.60\textwidth]{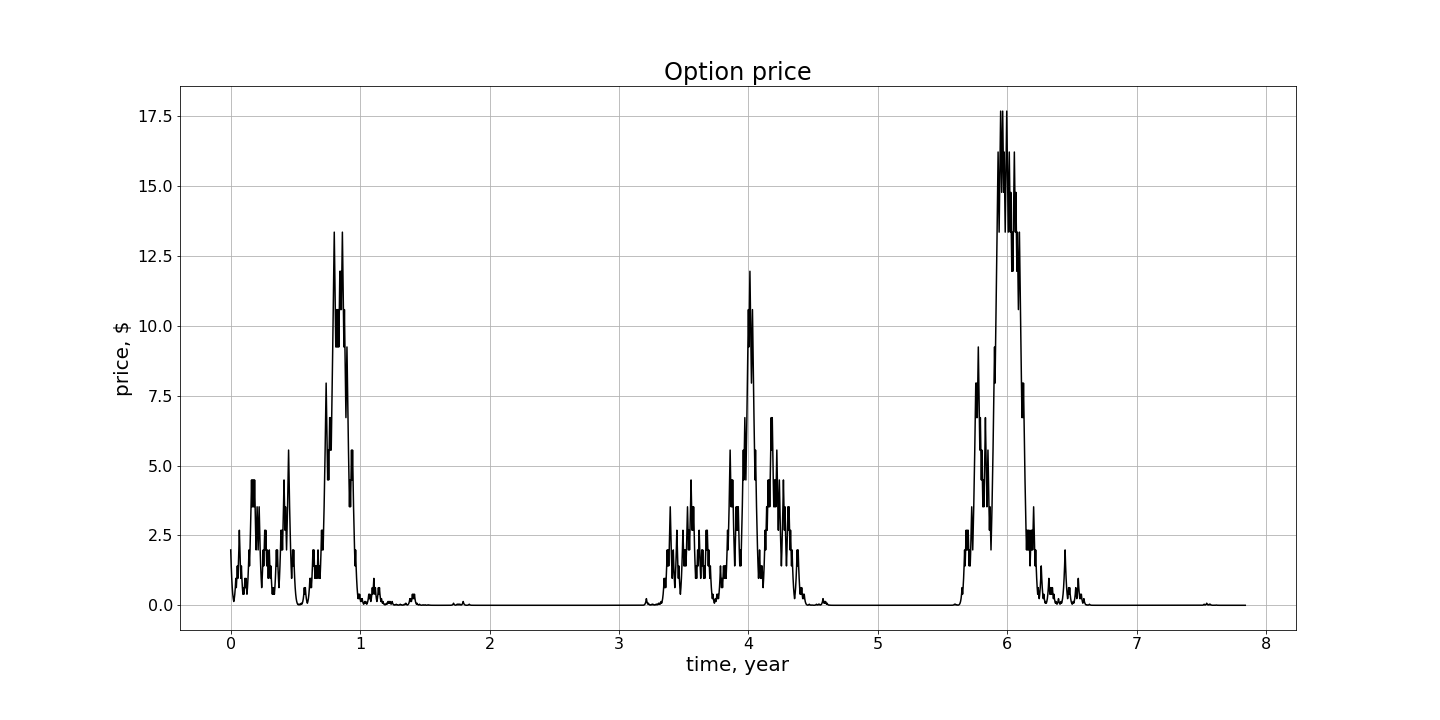}}
\par
\end{center}
\caption{\emph{\ The trajectory of the option price with }$\hat{\protect%
\sigma}=0.1$\emph{\ generated by the Black-Scholes formula (\protect\ref{2.2}%
) for the stock price trajectory of Figure 1. In this case, }$\protect\sigma 
$\emph{\ should be replaced with }$\hat{\protect\sigma}$\emph{\ in (\protect
\ref{2.2}). }}
\label{fig: The trajectory of the option price with}
\end{figure}

\subsection{Results}

\label{sec:6.3}

\begin{figure}[h]
\par
\begin{center}
{\includegraphics[width = 1.1\textwidth, height =
0.60\textwidth]{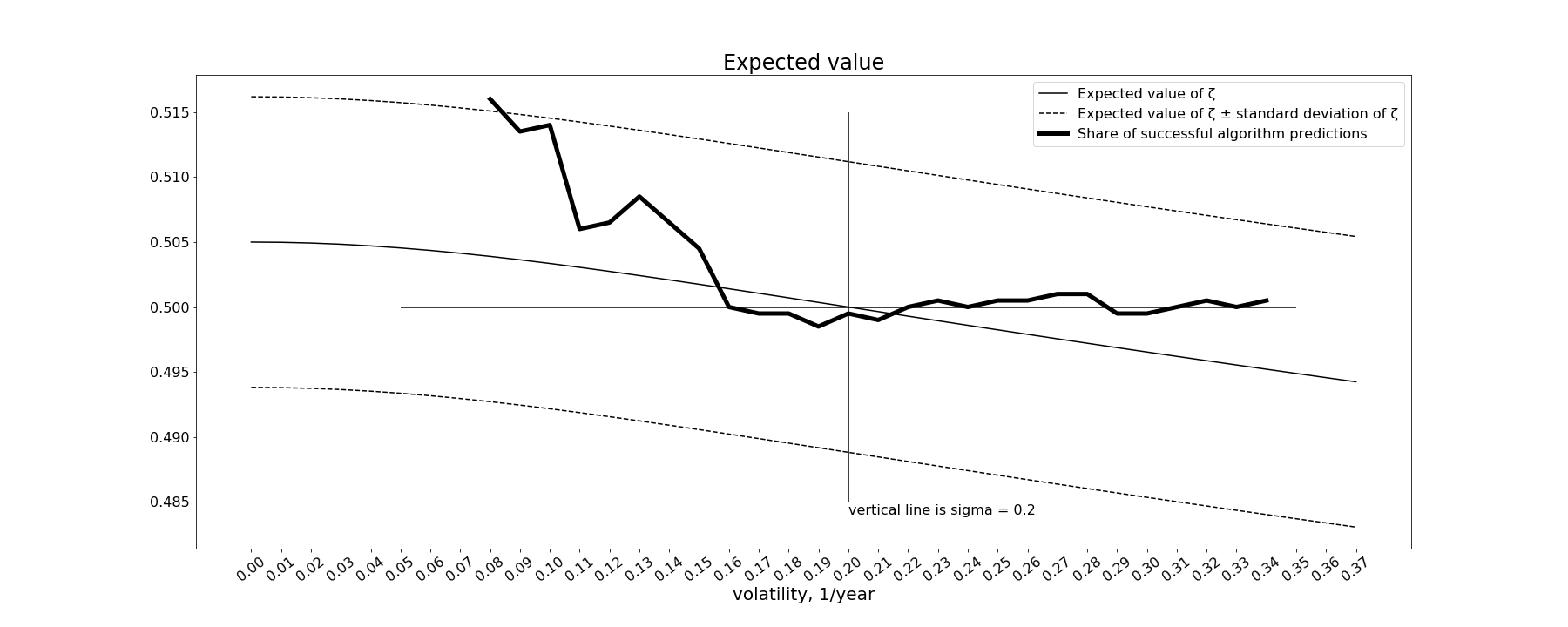}} 
\end{center}
\caption{ \emph{Results of our computations. The vertical line indicates the
value }$\protect\sigma =0.2$\emph{\ of the volatility of the stock in our
numerical studies. The middle curve corresponds to the ideal case when both
volatilities }$\protect\sigma $\emph{\ and }$\hat{\protect\sigma}$\emph{\
are known. This curve depicts the dependence of the mathematical expectation 
}$p\left( \hat{\protect\sigma}\right) $\emph{\ of the random variable }$%
\protect\zeta $\emph{\ on the market's opinion about the volatility }$\hat{%
\protect\sigma}$\emph{\ of the option, see (\protect\ref{4.1}). Two curves,
which are parallel to the middle one, are shifts of the latter by }$\protect%
\sqrt{D},$\emph{\ where }$D$\emph{\ is the dispersion of }$\protect\zeta ,$%
\emph{\ see (\protect\ref{4.01}). The bold faced curve represents the
frequency }$\overline{\protect\zeta }\left( \hat{\protect\sigma}\right) $%
\emph{\ of correctly predicted profitable cases for trading of this option
with the market's opinion }$\hat{\protect\sigma}$\emph{\ of the volatility
of the option. Our algorithm of section 3 was used to compute }$\overline{%
\protect\zeta }\left( \hat{\protect\sigma}\right) $\emph{. The meaning of }$%
\overline{\protect\zeta }\left( \hat{\protect\sigma}\right) $\emph{\ is
similar with the meaning of the ideal case of the random variable }$\protect%
\zeta .$\emph{\ Thus, it is natural that the bold faced curve lies in the
trust corridor of }$\protect\zeta .$ }
\label{fig:Results of our computations}
\end{figure}

For every discrete value of $\hat{\sigma}$ in (\ref{6.1}), we introduce the
sequence $\left\{ \overline{\xi }_{j}\left( \hat{\sigma}\right) \right\}
_{j=1}^{n}.$ This sequence is similar with the sequence $\left\{ \xi
_{j}\right\} _{j=1}^{n}$ in (\ref{3.99}). Recall that $v(s(t_{j}),t_{j})$
and $v(s(t_{j}+y),t_{j}+y)$ are true prices of the option at the moments of
time $t_{j}$ and $t_{j}+y$ respectively. We set%
\begin{equation}
\overline{\xi }_{j}\left( \hat{\sigma}\right) =\left\{ 
\begin{array}{c}
1\text{ if }v_{\text{pred}}\left( t_{j}+y\right) \geq v(s(t_{j}),t_{j})\text{
and }v(s(t_{j}+y),t_{j}+y)\geq v(s(t_{j}),t_{j}), \\ 
0\text{ otherwise.}%
\end{array}%
\right.  \label{6.3}
\end{equation}%
Next, we introduce the function $\overline{\zeta }\left( \hat{\sigma}\right) 
$ of the discrete variable $\hat{\sigma}$ as: 
\begin{equation}
\overline{\zeta }\left( \hat{\sigma}\right) =\frac{1}{n}\sum_{j=1}^{n}%
\overline{\xi }_{j}\left( \hat{\sigma}\right) ,\text{ }n=2000.  \label{6.4}
\end{equation}%
It follows from (\ref{6.3}) and (\ref{6.4}) that $\overline{\zeta }\left( 
\hat{\sigma}\right) $ is the frequency of correctly predicted profitable
cases for trading of this option with the market's opinion $\hat{\sigma}$ of
the volatility of the option. Predictions are performed by our algorithm of
section 3. Comparison of (\ref{3.99}) and (\ref{4.0}) with (\ref{6.3}) and (%
\ref{6.4}) shows that $\overline{\zeta }\left( \hat{\sigma}\right) $ is
similar with the ideal case of the random variable $\zeta .$ The bold faced
curve on Figure ~\ref{fig:Results of our computations} depicts the graph of
the function $\overline{\zeta }\left( \hat{\sigma}\right) .$ The middle
non-horizontal curve on Figure \ref{fig:Results of our computations} depicts
the graph of the function $p\left( \hat{\sigma}\right) ,$ which was
constructed in subsection 6.2 for the ideal case. The upper and the lowest
curves on Figure ~\ref{fig:Results of our computations} display the shifts
of the ideal curve up and down by $\sqrt{D},$ where $D$ is the dispersion of 
$\zeta $ and $D$ is given in (\ref{4.01}). In other words, there is a high
probability chance that the values of $\zeta $ are contained in the trust
corridor between these two curves. The vertical line indicates the
\textquotedblleft critical" value $\hat{\sigma}=\sigma =0.2,$ where $\sigma
=0.2$ is the volatility of the stock.

One can see from the bold faced curve of Figure ~\ref{fig:Results of our
computations} that as long as $\hat{\sigma}$ is either rather close to $%
\sigma $ or $\hat{\sigma}<\sigma ,$ the short position of this option
represents a significant risk. However, when $\hat{\sigma}$ becomes less
than $\approx 0.7\sigma ,$ the probability of the profit in the short
position increases. This coincides with the prediction of our theory.

The bold faced curve is an analog of the middle curve of the mathematical
expectation of the random variable $\zeta $ in the ideal case. Since the
bold faced curve on Figure \ref{fig:Results of our computations} lies within
the trust corridor of the ideal algorithm, then we conclude that our
prediction accuracy of profitable cases is comparable with the ideal one.

Unlike the ideal case, in a realistic scenario of the financial market data
of, e.g. \cite{Bloom} only the information about the approximate values of $%
\hat{\sigma}$ is available. It is this information, which was used in \cite%
{KlibGol,Nik} and, in particular, in Tables 1,2.

Thus, our results support the following trading strategy in the non-ideal
case:

\textbf{Trading Strategy for the Non-Ideal Case:}

Let $\eta >0$ be a threshold number, which should be determined numerically
by trial and error.\ For example, $\eta $ might probably be linked with the
transaction cost. Let $v_{\text{pred}}\left( t_{j}+y\right) $ be the number
defined in (\ref{6.2}).

\begin{enumerate}
\item If $v_{\text{pred}}\left( t_{j}+y\right) \geq v\left( s\left(
t_{j}\right) ,t_{j}\right) +\eta ,$ then it is recommended to buy the option
at the current trading day $t_{j}$ and sell it on the next trading $t_{j}+y.$

\item If $v_{\text{pred}}\left( t_{j}+y\right) <v\left( s\left( t_{j}\right)
,t_{j}\right) -\eta ,$ then it is recommended to go short at the current
trading day $t_{j}$, and to follow by closing the short position at the
trading day $t_{j}+y$.

\item If $v\left( s\left( t_{j}\right) ,t_{j}\right) -\eta \leq v_{\text{pred%
}}\left( t_{j}+y\right) <v\left( s\left( t_{j}\right) ,t_{j}\right) +\eta ,$
then it is recommended to refrain from trading.
\end{enumerate}

We believe that our results support the following two hypotheses:

\textbf{Hyphothesis 1:} \emph{The reason why the heuristic algorithm of \cite%
{KlibGol} and section 3 performs well is that it likely forecasts in many
cases the signs of the differences }$\sigma -\hat{\sigma}$\emph{\ for the
next trading day ahead of the current one.}

\textbf{Hyphothesis 2:} Since the maximal value of $\overline{\zeta }\left( 
\hat{\sigma}\right) $ in the bold faced curve of Figure ~\ref{fig:Results of
our computations} is 0.515, which is rather close to the value of 0.5577 in
the \textquotedblleft Precision" column of Table 1 and in the second column
of Table 2, then we probably had in those tested real market data about 56\%
of options, in which $\sigma -\hat{\sigma}<0.$

\section{Concluding Remarks}

\label{sec:7}

We have considered a mathematical model, in which two markets are in place:
the stock market and the options market. We have assumed that the market is
imperfect. More precisely, we have assumed that agents of the option market
have their own idea about the volatility $\hat{\sigma}$ of the option, and
this idea might be different from the volatility $\sigma $ of the stock. We
have proven that if that $\sigma \neq \hat{\sigma}$, then there is an
opportunity for a winning strategy. A rigorous probabilistic analysis was
carried out. This analysis has shown that the mathematical expectation of
the correctly guessed option price movements can be obtained, and it depends
on the difference between $\sigma $ and $\hat{\sigma}$.

We have considered both ideal and non-ideal cases. In the ideal case, both
volatilities $\sigma $ and $\hat{\sigma}$ are known. In the more realistic
non-ideal case, however, only the volatility $\hat{\sigma}\approx \sigma _{%
\text{impl}}$ of the option is known from the market data, see, e.g. \cite%
{Bloom}. We have demonstrated in our numerical simulations that the accuracy
of our prediction of profitable cases by the algorithm of \cite{KlibGol} for
the non-ideal case is comparable with that accuracy for the ideal case.

These results led us to two hypotheses. The first hypothesis is that our
algorithm of \cite{KlibGol} actually forecasts in many cases the signs of
the differences $\sigma -\hat{\sigma}$ for the next trading day ahead of the
current one. Our second hypothesis is based on our above results as well as
on the \textquotedblleft Precision" column in Table 1 and the second column
in Table 2. This second hyphothesis tells one that probably about 56\% out
of tested 23,549 options of \cite{Nik} with the real market data  had $%
\sigma -\hat{\sigma}<0.$

A new convergence analysis of our algorithm was carried out. To do this, the
technique of \cite{KlibYag} was modified and simplified for our specific
case of the 1-D parabolic equation with the reversed time. We have lifted
here the assumption of \cite{KlibGol} that the time interval $\left(
0,2y\right) $ is sufficiently small. Indeed, even though we actually work
with a small number $2y$ in our computations, that assumption might require
even smaller values. In addition, we have derived a stability estimate for
Problem 3 of subsection 4.1, which was not done in \cite{KlibGol}.

\begin{center}
\textbf{Acknowledgment}
\end{center}

The work of A.A. Shananin was supported by the Russian Foundation for Basic
Research, grant number 20-57-53002.

\end{document}